\documentclass[
reprint,
nofootinbib,
amsmath,amssymb,amsbsy,
prd,
onecolumn,
superscriptaddress,
longbibliography,
preprintnumbers]{revtex4-1}
\usepackage{lineno}
\usepackage[dvipsnames,usenames]{xcolor}
\usepackage{graphicx}
\usepackage{graphicx}
\usepackage{mathtools}
\usepackage[section]{placeins}
\usepackage{dcolumn}% Align table columns on decimal point
\usepackage{bm}% bold math
\usepackage{mathrsfs,natbib}
\usepackage[title]{appendix}
\usepackage{graphicx,epsf,amssymb,amsbsy,amsfonts,amssymb,amsmath}
\usepackage[colorlinks=true]{hyperref}
\hypersetup{
    unicode=false,          % non-Latin characters
    pdftoolbar=true,        % show Acrobat
    pdfmenubar=true,        % show Acrobat
    pdffitwindow=false,     % window fit to page when opened
    pdfstartview={FitH},    % fits the width of the page to the window
    pdftitle={},    % title
    pdfauthor={},     % author
    pdfsubject={},   % subject of the document
    pdfcreator={},   % creator of the document
    pdfproducer={}, % producer of the document
    pdfkeywords={} {} {}, % list of keywords
    pdfnewwindow=true,      % links in new window
    colorlinks=true,       % false: boxed links; true: colored links
    linkcolor=magenta, %red,          % color of internal links (change box color with linkbordercolor)
    citecolor=blue,        % color of links to bibliography
    filecolor=magenta,      % color of file links
    urlcolor=blue           % color of external links
}
\usepackage{slashed}
\usepackage{comment}
\usepackage[dvipsnames]{xcolor}
\usepackage{color,soul}
\usepackage{simplewick}

\newcommand{\be}{\begin{equation}}
\newcommand{\ee}{\end{equation}}

\newcommand{\mybar}[1]

\newcommand{\bea}{\begin{eqnarray}}
\newcommand{\eea}{\end{eqnarray}}
\newcommand{\nn}{\nonumber}

\newlength{\backup}

\begin{document}
%\linenumbers

\title{Viability of perturbative expansion for quantum field theories on neurons}
\author{Srimoyee Sen}%
\email{srimoyee08@gmail.com}
\affiliation{Department of Physics and Astronomy, Iowa State University, Ames, Iowa 50011, USA}%
\author{Varun Vaidya}%
\email{Varun.Vaidya@usd.edu}
\affiliation{Department of Physics, University of South Dakota, Vermillion, SD 57069, USA}%

\date{\today}

\begin{abstract}
Neural Network (NN) architectures that break statistical independence of parameters have been proposed as a new approach for simulating local quantum field theories (QFTs)  \cite{Demirtas:2023fir}. In the infinite neuron number limit, single-layer NNs can exactly reproduce QFT results. This paper examines the viability of this architecture for perturbative calculations of local QFTs for finite neuron number $N$ using scalar $\phi^4$ theory in $d$ Euclidean dimensions as an example. We find that the renormalized $O(1/N)$ corrections to two- and four-point correlators yield  perturbative series which are sensitive to the UV cut-off and therefore have a weak convergence.  We propose a modification to the architecture to improve this convergence and discuss constraints on the parameters of the theory and the scaling of N which allow us to extract accurate field theory results.

\end{abstract}
\maketitle

\section{Introduction}

Accelerated progress in machine learning (ML) over the past decade has had significant impact across many research domains, including physics, and has motivated substantial interdisciplinary work. At the intersection of physics and machine learning, two prominent practical questions have emerged:

\begin{enumerate}
    \item Can techniques from statistical mechanics and the path integral formulation of quantum field theory (QFT) help us build a theoretical understanding of how neural networks learn?
    \item Can neural networks be used to facilitate computations in quantum field theory?
\end{enumerate}

These two questions are deeply interrelated, and will motivate the questions we explore in this work. The second question itself splits naturally into two subcategories: (a) applied machine learning for physics problems \footnote{The area of applied ML to physics has already seen considerable progress. Examples include pattern recognition in large datasets \cite{Duperrin:2023elp, ATLAS:2023ber, ATLAS:2023ixc, Gambhir:2022gua, Gambhir:2022dut},  materials synthesis in condensed matter physics \cite{material:ai, osti_1211231, article, article2, article3, PhysRevMaterials.9.053803}, evaluating lattice QFT path integrals using normalizing flows \cite{PhysRevD.100.034515, PhysRevD.106.074506}, contour deformation \cite{Detmold:2021ulb, RevModPhys.94.015006, Alexandru_2017} and more \cite{Gerdes_2023}.} and (b) the theoretical interplay between machine learning and QFT techniques.
In this paper we concentrate on the theoretical connection between ML and QFT. Several works in the past decade have noted a correspondence between neural networks and Gaussian processes \cite{matthews2018gaussianprocessbehaviourwide, novak2020bayesiandeepconvolutionalnetworks,garrigaalonso2019deepconvolutionalnetworksshallow,
yang2020scalinglimitswideneural,
yang2021tensorprogramsiwide,
yang2020tensorprogramsiineural}, providing insights into the learning dynamics of neural networks. Furthermore, this correspondence has inspired a novel paradigm for computing path integrals in QFT using neural network architectures \cite{Halverson:2021aot,Demirtas:2023fir}. See \cite{Ferko:2025ogz, Halverson:2024axc, Howard:2024kfd} for related work. The corresponding formulation of QFT on neural networks is known as neural network field theory or NNFT and this is the subject of this paper. Path integrals calculated using this framework have the potential to be useful in extracting non-perturbative physics from quantum field theories of interest. And if successful, this technique may prove complementary to other non-perturbative techniques including lattice field theory. Before such calculations are implemented in practice, it is important to understand how renormalization works in NNFT, which observables can be easily extracted using this framework and given that any computer simulation will involve finite resources, how errors scale as a function of these resources. The aim of this paper is to address some of these questions by critically examining the NNFT paradigm within the context of a simple relativistic scalar field theory: the $\phi^4$ theory.\\% for a finite width neural network where width stands for the number of neurons.\\

NNFT formulation of QFT path integrals is based on a single layer neural network of width $N$.
A key ingredient of the NNFT architecture is its activation function. One of the earliest papers on the subject, \cite{Halverson:2021aot} demonstrated how one can build relativistic free scalar field theory path integrals by appropriately designing activation functions. The resulting correlation functions match free quantum field theory results in the infinite-width (neuron number, $N \to \infty$) limit. Of course, any real NNFT simulation will involve finite $N$. Moreover, any nontrivial application of this framework will have to include interactions. This brings one to the natural follow up
%The goal of this paper is twofold
\begin{itemize}
\item to introduce local interactions in the NNFT framework, 
\item to systematically estimate finite width ($N$) contributions to interacting theory correlation functions.
\end{itemize}
%The latter is motivated by limitations of a real simulation which inevitably involves finite (albeit large) $N$.
Finite width corrections have already been a subject of intense study for machine learning applications which has led to a new understanding of Neural Networks as Effective Field theories(EFTs) \cite{Roberts:2021fes,
Banta:2023kqe}. Additionally, it has been shown that for free field theories, the effective action of NNFT deviates from the target QFT by terms suppressed as $1/N$. These corrections are entirely non-local and can be thought of as operators inducing non-local interactions in the effective theory \cite{Erbin:2021kqf} \footnote{One may be tempted to utilize such non-Gaussian contributions to model interacting field theories relevant for physical systems. However,
QFTs describing physical systems in nature, typically have \textit{local} interactions. Thus the non-Gaussian finite width corrections of NNFT cannot be used for describing such local interactions.}. In the context of NNFT, it was proposed that the desired local interactions can be incorporated within the framework \cite{Demirtas:2023fir} by breaking statistical independence of network parameters, i.e. biases and weights. Since then, there also has been considerable work in extending QFT formulations on neural networks for fermions and supersymmetry\cite{Huang:2025ipy,Frank:2025zuk}, String theory (\cite{Robinson:2025ybg,Frank:2026bui} and most recently  2d Liouville theory \cite{Ferko:2026axm} which also explores the paradigm of realizing field theories exactly in a finite width network. \\ %Our focus will be to quantify the deviations of NNFT correlation functions at finite width from their quantum field theoretic counterparts, and to assess the viability of NNFTs as computational tools for local quantum field theories. 

In this paper, we  focus on the utility of the NNFT for doing perturbative calculations in the local interactions of a relativistic QFT. This is important for several reasons.  A perturbative calculation provides an essential connection with Monte Carlo simulations in the regime of weak coupling. It also provides us with an analytical understanding of the finite width errors and therefore the possibility of improving the NN architecture to reduce the errors rather than simply increasing the width which may not be cost effective. Most importantly, Monte Carlo simulations and perturbation theory must agree at weak couplings. This agreement will give us an indication of the reliability of future Monte Carlo simulation in the non-perturbative regime. To this end, we aim to quantify the neuron number $N$ required at a given perturbative order to ensure that finite
$N$ errors are under control. To understand why smallness of finite $N$ correction is not a foregone conclusion, 
note that, QFT path integrals include several dimensionless scales, e.g. the ratio of correlation length($\xi$) to inverse UV  momentum cutoff ($\Lambda^{-1}$). Similarly, a finite space-time volume will give rise to another dimensionless scale,  $V\Lambda^4$. While the errors associated with NNFT when compared with the QFT are expected to go as inverse power of $N$, the scaling of the error with these other dimensionless ratios is not generally known. Whether the errors get enhanced or suppressed by these dimensionless quantities can only be revealed through an explicit calculation. This relates to the final question addressed in this paper, that of renormalizability. Even for weakly interacting QFT path integral, perturbative expansions in bare parameters of the theory are often plagued with large corrections that scale with the UV cut-off of the theory leading to poor convergence. However re-organizing the calculation in terms of renormalized parameters significantly improves the convergence and hence reliability of perturbation theory. So, the natural question is to what extent can finite N corrections be dealt with in the same manner in NNFT.  

To address these questions concretely, we compare relativistic real scalar field theory with its NN version (NNFT). We analyze connected correlation functions of free relativistic real scalar field theory, as well as that of an interacting $\phi^4$ theory at finite $N$. The interacting theory analysis is performed perturbatively in the quartic coupling at one loop. 

We show that the error in prediction after renormalization in momentum space can be minimized by avoiding certain special points in phase space where the errors are enhanced by the powers of volume measured in units of the inverse momentum cut-off ($\Lambda^{-1}$), i.e. $V\Lambda^4$. Computing observables at these special points can be expensive due to the requirement of $N$ having to be large enough to compensate for this enhancement. 
In the rest of the phase space, where such enhancements by space-time volume are absent, we find that the tree level and one-loop renormalized NNFT observables experience $1/N$ errors that can be enhanced by powers of a smaller dimensionless ratio $\Lambda \xi$. As a result, a direct implementation of the proposal in \cite{Demirtas:2023fir} leads to a weakly convergent perturbative series at $O(1/N)$. We propose a modification of the NN architecture to improve this and discuss the resulting constraints on the network width N. 

This paper is organized as follows: 
We begin by reviewing the proposal utilizing NN architectures for simulating QFT in Section \ref{sec:NNQFT}. We compute correlation functions for the free scalar field theory at finite N and analyze the scaling of the leading order(1/N) finite width errors in Section \ref{sec:FreeQFT}. The corresponding results for interacting $\phi^4$ theory to one loop are presented in Section \ref{sec:IntQFT}. We perform one loop perturbative renormalization for the NNFT in Section \ref{sec:Renorm} for the two and four point correlators and propose a modified framework in \ref{sec:IPT} to improve the predictions. We then comment on the results at higher order in perturbation theory and higher point correlators in Section \ref{sec:HighPT}. We summarize our results and speculate on possible improvements in the NN architecture in Section \ref{sec:Summary}.

\section{NN for simulating QFT}
\label{sec:NNQFT}
Neural networks are the building blocks of machine learning (ML). A typical ML implementation involves processing training data using a neural network architecture, the output of which is then compared to the desired output to facilitate learning. A neural network architecture can be composed of several layers, each of which except the last contain several neurons, neuron number denoted as $N$. The last layer usually contains one neuron known as the output neuron. 
The number of layers denotes the depth of the network and the number of neurons for each layer is referred to as the width of that layer. Every layer takes in data which we can represent as $x_j$ with the subscript $j$ takes values in $1, \cdots d$ where $d$ is the dimensionality of the data. A neuron in any layer layer first computes a linear function of the input $x_j$ 
\bea
z_i(x)=b_{ij}x_j + c_i
\eea 
where $i$ denotes neuron number, $b_{ij}$ are called weights and $c_i$ are biases. $b$ and $c$ are parameters of the NN which are drawn from some distribution functions that are relevant for the particular application in question. 
The architecture of each layer is decided  by an activation function  $\sigma$ which acts on $z_i$ and produces the output of the layer. E.g. the neuron $i$ could produce an output of $1$ if $z_i$ is greater than some threshold and output zero if $z_i$ if it is smaller than the threshold. This is the simplest activation function and the binary output of the neuron is at times referred to as firing of the neuron. However, activation functions can be any function depending on the goal of the network. 

The basis for utilizing NN for simulating Quantum field theory lies in the NN/ Gaussian Process (NNGP)  correspondence. For most modern NN architectures, it can be shown using the Central Limit Theorem(CLT) that  for a single layer NN, in the limit of the NN layer width $ N \rightarrow \infty$, the NN is a draw from a Gaussian Process(GP) \cite{matthews2018gaussianprocessbehaviourwide, novak2020bayesiandeepconvolutionalnetworks,garrigaalonso2019deepconvolutionalnetworksshallow,yang2020scalinglimitswideneural,yang2021tensorprogramsiwide,yang2020tensorprogramsiineural} . This establishes a connection with free field theories which are also Gaussian and therefore can be realized in an NN architecture. A finite width N leads to non-Gaussianities that are equivalent to non-local interactions of the field theory.

This correspondence is best illustrated with an example, a real scalar field theory in $d$ Euclidean space-time dimensions. A free scalar field theory in d Euclidean dimensions is described by the action 
\bea
S_E= -\frac{1}{2}\int d^dx  \phi(x) \left( \nabla^2+ m^2 \right)\phi(x)
\eea
where m is the mass. The partition function for this field theory is 
\bea
Z_J = \int D\phi(x) e^{- S_E+ \int d^dx \phi(x) J(x)}
\label{eq:FPart}
\eea
All observables in this theory can be computed through correlation functions defined as 
\bea 
\langle \phi(x_1) ... \phi(x_n) \rangle = \int D\phi(x) \phi(x_1) ... \phi(x_n)e^{-S_E} = \frac{\delta^n}{\delta J(x_1) ..\delta J(x_n)} Z_J\Big|_{J=0}
\eea
Our interest therefore is to compute these correlation functions by exploiting the NNGP correspondence. 

The computation of these correlators on the NN is realized as follows. 
The NNFT is implemented on a single hidden layer NN with N neurons and an output layer with a single neuron. The width of the first or input layer is N and the activation function for this layer is chosen to be the cosine function, so that the output of a neuron i in the first layer is $\cos(z_i) = \cos(\sum_{j=1}^d b_{ij}x_j+c_i)$. These N outputs are then fed to the output layer which has a single neuron and produces a single number $\phi(x)$ as the final result of the NN. 
\bea
\phi(x)= \left(\frac{\sqrt{2 V_d}}{\sigma_a(2\pi)^{d/2}}\right)\sum_{k=1}^N \frac{a_k \cos(b_{kj}x_j+c_k)}{\sqrt{\sum_{j=1}^db_{kj}^2+m^2}}
\label{eq:Act}
\eea
where repeated indices are summed over. 
The output $\phi(x)$ will play the role of the scalar field and $m$ is going to play the role of the scalar field mass. This particular choice for the activation function was motivated in \cite{Halverson:2021aot,Demirtas:2023fir} to reproduce the results for free scalar field theory as we explain below.  $a$, $b$ and $c$ are parameters of this neural network. $a_k$ are drawn from a Gaussian distribution with zero mean. $b_{kj}$ and $c_i$ are drawn from a uniform distribution with $b_{ij}^2\leq \Lambda^2$ where $\Lambda$ will play the role of the momentum cutoff of the theory. This can be phrased as drawing $b$ from a d dimensional sphere of radius $\Lambda$ with uniform probability. We denote the corresponding distribution function $P_{b_{ij}}$. The volume of this sphere is denoted as $V_d$. Similarly, $c_k$ are phase factors that are uniformly drawn from the interval $-\pi<c_i\leq \pi$ and we denote the corresponding distribution function as $P_c$. $P_a$, $P_{b_{ij}}$ and $P_c$ can be expressed as
\bea 
P_{a_i}&=& \frac{\sqrt{N}}{\sqrt{2\pi}\sigma_a}e^{-N a_i^2/2\sigma_a^2}\nonumber\\
P_{b_{ij}}&=&\frac{\Theta\left(\Lambda-\sqrt{\sum_j b_{ij}^2}\right)}{V_d}\nonumber\\
P_{c_i}&=&\frac{\left(1-\Theta((c_i+\pi)(c_i-\pi))\right)}{2\pi}
\eea
where $\Theta$ is the unit step function. The corresponding NNFT partition function is given by 
\bea
Z_{\text{Free}}^{\text{NN}}(J)=\int db_{ij} da_i dc_i P_a P_{b_{ij}} P_c e^{\int d^dx J(x) \phi(x)}
\eea 
This is called the \textit{parameter space representation} which can be thought of as a dual description of the field theory partition function in Eq.~\ref{eq:FPart}. 
The $n$-point correlation function $\langle\phi(x)\phi(y)\phi(z)..\rangle$ is computed by computing the ensemble averages of the product $\phi(x)\phi(y)\phi(z)..$ for distinct d dimensional inputs $x,y, z ..$ over the parameter values of $a, b$ and $c$. Symbolically, this can be expressed as
\bea
\langle\phi(x)\cdots \phi(w) \rangle = \prod_{i=1}^N \int db_{i1}... \int b_{id} \int da_i \int \frac{dc_i}{2\pi}\,\, \phi(x)\cdots \phi(w) P_{a_i} P_{b_{ij}} P_{c_i}
\label{nnpart}
\eea 
with the field $\phi(x)$ defined in terms of the NN parameters (Eq.~\ref{eq:Act}). From here on we will write $\int db_{i1}... \int b_{id} = \int d^db_i$ since these parameters will play the role of the d dimensional momentum integral. 
As we will see in the next section, the two-point correlator computed using the activation function of Eq. \ref{eq:Act} reproduces the free scalar field theory correlation function exactly. Higher point correlators for the free theory are expected to match the QFT result up to $1/N$ corrections. However, our ultimate interest is in simulating local interacting field theories which describe most of the physical systems of interest.
A classic example which we will be the main focus of this paper is $\phi^4$ theory which is described by the field theory action 
\bea
S= S_E - \frac{\lambda}{4!}\int d^dx\,\,\phi^4(x) 
\eea
where $\lambda$ is the coupling strength. In this paper, we are going to focus on theories in $d\leq 4$ dimensions. Note that, $\phi^4$ theory is quantum trivial in $d=4$ which in general implies that we cannot take the cutoff to infinity while retaining an interacting theory in the infrared. Our goal here is to merely assess the renormalizability of the low energy EFT with a finite cut-off which is well defined. Thus, we include $d=4$ in our analysis.

We note that any formulation of NNFT at finite width $N$ automatically induce some non-local $1/N$ interactions. However, these cannot be utilized to construct local interactions of the form we are interested in. 
A possible pathway to engineering this interaction in NNFT  was suggested in \cite{Demirtas:2023fir} which proposes to introduce local interactions by breaking the statistical independence of network parameters. Local interacting field theories can be incorporated in this construction by simply modifying the probability distribution for network parameters by the local interaction term written as a function of $\phi(x)$. E.g., for a $\phi^4$ theory, one could sample $a_i, b_{ij}, c_i$ from a joint distribution of the form
\bea
P_{abc}\propto\prod_{i=1}^N P_{a_i}P_{b_{ij}} P_{c_i} e^{-\frac{\lambda}{4!}\int d^d x\phi^4(x)}.
\label{eq:Pfree}
\eea
where $\phi(x)$ is again given by Eq.~\ref{eq:Act}.
 In the above expression repeated indices are not automatically summed. Instead we have chosen to write the sums explicitly wherever needed. 
 We can then define an interacting NN partition function of the form

\bea
Z^{\text{NN}}_J=\int d^db_{i} da_i dc_i P_{abc}(J)
\eea
where  
\bea
P_{abc}(J)\propto\prod_{i=1}^Ne^{-\frac{N}{2\sigma_a^2} a_i a_i-\frac{\lambda}{4!}\int d^d x\left(\phi^4(x)-J(x) \phi(x)\right)}\left(1-\Theta((c_i+\pi)(c_i-\pi))\right)\Theta\left(\Lambda-\sqrt{\sum_j b_{ij}^2}\right).
\eea
and J is the source term for the scalar field.
The $n$-point correlator computed using normalized distribution functions can then be expressed as 
\bea
\langle\phi(x_1)..\phi(x_n)\rangle=\frac{\partial^n \log Z_J^{\text{NN}}}{\partial J(x_1)..\partial J(x_n)}\bigg|_{J=0}
\label{eq:Corr}
\eea
In the next few sections, we will now compute correlation functions in NNFT for both the free and interacting theory. We will focus on the two and four point correlators and discuss perturbative renormalization with the aim of understanding the errors induced by the finite network width.

%%%%%%%%%%%%%%%%%%%%%%%%%%%%%%%%%%%%%%%%%%%%%%%%%%%%%%%%%%%%%%%%%%%%%%%%%%%%%%%%%%%%%%%%%%%%%%%%%%

\section{Free field theory at finite N}
\label{sec:FreeQFT}
\subsection{Two point correlator}
\label{sec:2pFT}
 
We start off by computing the two point correlator in NNFT. As shown in \cite{Demirtas:2023fir} for a single hidden layer NN, the architecture developed in the previous section should reproduce this exactly and there should not be any $1/N$ corrections. Using Eq.~\ref{eq:Act}  in Eq.~\ref{nnpart} we can write the correlation function as 
\bea
\langle \phi(x) \phi(y) \rangle &=&  \left(\frac{2 V_d}{\sigma_a^2(2\pi)^d}\right)\prod_{i=1}^N  \Bigg[ \int da_i  \frac{\sqrt{N}}{\sqrt{2\pi} \sigma_a} e^{-\frac{N}{2\sigma_a^2}a_ia_i}\int_{V_d} \frac{d^db_{i}}{V_d}\int_{-\pi}^{\pi} \frac{dc_i}{2\pi} \Bigg] \nn\\
&& \left( \sum_k^{N} \frac{a_i \cos (b_k \cdot x+c_k)}{\sqrt{b_k^2+m^2}}\right) \left( \sum_m^{N} \frac{a_m \cos (b_{m}\cdot y+c_m)}{\sqrt{b_m^2+m^2}}\right).\nonumber\\
\label{e1}
\eea
where we will use the notation $b_k \cdot x= \sum_{j} b_{kj}x_j$
Performing the integral over $a_i$ followed by $c_i$, we obtain 

\bea
G_c^{(2)}(x,y)= \langle \phi(x) \phi(y) \rangle &=&\int_{V_d} \frac{d^db}{(2\pi)^d} \left(  \frac{\cos (b \cdot (x-y))}{b^2+m^2}\right)=  \int_{V_d} \frac{d^db}{(2\pi)^d}  \frac{e^{i b \cdot(x-y)}}{b^2+m^2}
\label{e2}
\eea
which gives the correctly normalized two point correlator.  We see that the result is independent of N and therefore receives no $1/N$ corrections. 
There is still however a cutoff $\Lambda$ on the b integral. Since we are looking to probe length scales of O(1/m) or longer, we need $\Lambda \gg m$ for the error to be small. Ideally we would set $\Lambda$ to the scale of new physics. 
%%%%%%%%%%%%%%%%%%%%%%%%%%%%%%%%%%%%%%%%%%%%%%%%%%%%%%%%%%%%%%%%%%%%%%%%%%%%%%%%%%%%%%%%%%%%%%%%%%%%%%%%%%%%%%%%%%%%%%%
\subsection{Four point correlator}

In the same manner we can compute the 4 point correlator. In the free QFT, the connected 4 point correlator should be zero, but in NNFT the finite width will introduce $1/N$ non-Gaussian corrections 
\bea
&&\langle \phi(w_1) \phi(w_2)\phi(w_3) \phi(w_4) \rangle \rangle \equiv G^{(4)}(w_1,w_2,w_3,w_4)\nonumber\\
&&=  \left(\frac{2 V_d}{\sigma_a^2(2\pi)^d}\right)^2 \prod_{i=1}^N  \Bigg[ \int da_i  \frac{\sqrt{N}}{\sqrt{2\pi} \sigma_a} e^{-\frac{N}{2\sigma_a^2}a_ia_i}\int_{V_d} \frac{d^db_{i}}{V_d}\int_{-\pi}^{\pi} \frac{dc_i}{2\pi} \Bigg]\nn\\
&& \sum_k^{N} \frac{a_k \cos (b_{k} \cdot w_1+c_k)}{\sqrt{b_k^2+m^2}} \sum_m^{N} \frac{a_m \cos (b_{m}\cdot  w_2+c_m)}{\sqrt{b_m^2+m^2}} \sum_o^{N} \frac{a_o \cos (b_{o}\cdot w_3+c_o)}{\sqrt{b_o^2+m^2}} \sum_q^{N} \frac{a_q \cos (b_{q} \cdot w_4+c_q)}{\sqrt{b_q^2+m^2}}\nonumber\\
\label{four}
\eea
 The connected 4 point correlator defined as 
 \bea
G_c^{(4)}(w_1,w_2,w_3,w_4) = G^{(4)}(w_1, w_2, w_3, w_4) - \Big\{G^{(2)}(w_1, w_2)  G^{(2)}(w_3, w_4) + \text{permutations}\Big\}\eea
The details of this calculation are given in Appendix \ref{App:4p} and gives
\bea
&&G_{c}^{(4)}(w_1,w_2,w_3,w_4)= \prod_{i=1}^4\Bigg[ \int_{V_d} \frac{d^db_{i}}{(2\pi)^d}   e^{ib_{i} \cdot w_i}\Bigg]\delta^d(b_{1}+b_{2}+b_{3}+b_{4}) \nn\\
 &&\Bigg[ \frac{1}{2N}\frac{V_d}{(2\pi)^d} \frac{\delta^d(b_1-b_2)\delta^d(b_1+b_3) + \delta^d(b_1+b_2)\delta^d(b_1-b_3)+\delta^d(b_1+b_2)\delta^d(b_1+b_3)}{(b_1^2+m^2)^2}\nn\\
 &-& \frac{1}{N} \frac{\delta^d(b_1+b_2)}{(b_1^2+m^2)(b_3^2+m^2)}-\frac{1}{N} \frac{\delta^d(b_1+b_3)}{(b_1^2+m^2)(b_2^2+m^2)}-\frac{1}{N} \frac{\delta^d(b_1+b_4)}{(b_1^2+m^2)(b_3^2+m^2)}\Bigg]
  \label{eq:fourv2}
 \eea
We see that the expression in \ref{eq:fourv2} has support only at specific external momentum configurations and is zero elsewhere. At these special kinematic points(SKP's), the correlator gets enhanced by factors of $V_s V_d$ where $V_s$ is the $d$ dimensional Euclidean space-time volume.
This will be important later when we look at renormalization. This 4 point correlator can be interpreted as arising out of an effective interaction induced by the non-Gaussianities in the free theory. We can estimate the strength of this interaction in momentum space by considering the ratio of this correlator with the tree level 4 point correlator, just as one would to estimate the coupling strength of a local $\phi^4$ interaction in a field theory
    \bea
    \lambda_p = \frac{\prod_{i=1}^4\Bigg[ \int d^dw_{i} e^{-ib_{i} \cdot w_i}\Bigg]G_{c}^{(4)}(w_1,w_2,w_3,w_4)}{\prod_{i=1}^4\Bigg[\frac{1}{b_i^2+m^2}\Bigg]\delta^d(b_{1}+b_{2}+b_{3}+b_{4})}.\nonumber\\
    \label{lp}
    \eea
For momenta that explicitly conserve total momentum 
\bea
\lambda_p &=&\Big[\prod_{i=1}^4(b_i^2+m^2)\Big]\Bigg[ \frac{1}{2N}\frac{V_d}{(2\pi)^d} \frac{\delta^d(b_1-b_2)\delta^d(b_1+b_3) + \delta^d(b_1+b_2)\delta^d(b_1-b_3)+\delta^d(b_1+b_2)\delta^d(b_1+b_3)}{(b_1^2+m^2)^2}\nn\\
 &-& \frac{1}{N} \frac{\delta^d(b_1+b_2)}{(b_1^2+m^2)(b_3^2+m^2)}-\frac{1}{N} \frac{\delta^d(b_1+b_3)}{(b_1^2+m^2)(b_2^2+m^2)}-\frac{1}{N} \frac{\delta^d(b_1+b_4)}{(b_1^2+m^2)(b_3^2+m^2)}\Bigg]
 \label{qe:lp2}
\eea

The scaling of these contributions can be estimated by looking at the point $b_i=0$, where
\bea
&&\lambda_p \sim \frac{1}{N}\Bigg[ \frac{3}{2}(V_s m^4) \Lambda^d    V_s -3V_s m^4 \Bigg] 
\nonumber\\
\label{lp3}
\eea
where we have used $V_d\sim \Lambda^d$. This would suggest that in, say, d=4, for finite N corrections to be small for computing correlators in momentum space, we need  $N \gg V_s m^4 (\Lambda^4 V_s)$, which can be too expensive. On the other hand if we only make predictions for momenta away from these SKP's in phase space, the error is zero. This feature can be exploited for extracting useful physics with small error. Note that, in a realistic numerical implementation in a finite volume, the SKP is likely to broaden into a special kinematic region (SKR) around the SKP, set by the corresponding momentum resolution. Thus, to avoid incurring order one errors in a numerical implementation, we will have to exclude the SKP as well as a finite width region around it. The width of this region can be systematically reduced by considering large volume.

We may also be interested in computing spatial correlators in which case, we can also define an effective interaction strength in position space using 
 \bea
    \lambda_x \equiv \frac{G_{c}^{(4)}(w_1,w_2,w_3,w_4)}{\prod_{i=1}^4\Bigg[ \int_{V_d} \frac{d^db_{i}}{(2\pi)^d}   \frac{e^{ib_{i} \cdot w_i}}{b_i^2+m^2}\Bigg]\delta^d(b_{1}+b_{2}+b_{3}+b_{4})}
    \label{eq:lx}
    \eea
    where we substitute $G_{c}^{(4)}(w_1,w_2,w_3,w_4)$ from Eq. \ref{eq:fourv2} in Eq. \ref{eq:lx}.
We immediately see that there is no singular behavior in position space. To get an estimate of the effective interaction strength, we observe that if we only probe  distances $w_i-w_j \sim 1/m \gg 1/\Lambda$, then the exponential factors $\exp{[i (w_i-w_j) b]}$ in Eq. \ref{eq:fourv2} put a cutoff of order $m$ on the momentum integrals. In this case, a simple dimensional analysis suggests that $ \lambda_x \sim m^{4-d}\frac{\Lambda^d}{N m^d}$. An interaction strength of 1 in d dimensions corresponds to $\lambda_x \sim m^{4-d}$, so that for 1/N correction to be small we need $N \gg \Lambda^d/m^d$. 
 This is much less severe than the enhancement in momentum space by a factor of $V_s V_d$ at the SKP's. We can understand this by noting that when we take the Fourier transform of the position space correlation function, the error in position space adds up coherently for specific momentum configurations but cancels out in most of the phase space. This also holds true for higher point correlators. It can be verified that the connected $2n$ point correlator starts off at O($1/N^{n-1}$)  but only has support at SKP's in momentum space. Likewise based on our calculation of the four- point correlator, we can see that the error for the $2n$ point correlator in position space will scale as $(\Lambda^d/(N m^d))^{n-1}$. Thus the condition $N \gg \Lambda^d/m^d$ is sufficient for small finite N errors in position space. On the other hand, for momentum space correlators, if we stay away from certain specific kinematic regions, the finite N error is zero, which is remarkable. In the next section we want to explore to what extent these conclusions hold for an interacting theory which is our main interest.

\section{Interacting theory at finite N}
\label{sec:IntQFT}

We now consider the implementation of the interacting theory on the finite width NN according to the prescription presented in section \ref{sec:NNQFT}. We will present a perturbative analysis to obtain an estimate of the leading (O(1/N)) error  after renormalization. 

We can now use the correspondence established between the field theory partition function implemented on the Neural Network in Eq.\ref{eq:Corr} to compute correlation functions. 
The $n$ point correlator in interacting theory is
\bea
\langle \phi(w_1) \phi(w_2) . . .\phi(w_n) \rangle =  \frac{\prod_{i=1}^N  \Bigg[ \int da_i  \frac{\sqrt{N}}{\sqrt{2\pi} \sigma_a} e^{-\frac{N}{2\sigma_a^2}a_ia_i}\Bigg]\Bigg[ \frac{1}{V_d}\int_{V_d} db^d_{i}  \Bigg] \Bigg[\frac{1}{2\pi} \int_{-\pi}^{\pi} dc_i \Bigg]e^{-\frac{\lambda}{4!}\int d^dx \phi^4(x)} \phi(w_1) \phi(w_2) . . .\phi(w_n)}{\prod_{i=1}^N  \Bigg[ \int da_i  \frac{\sqrt{N}}{\sqrt{2\pi} \sigma_a} e^{-\frac{N}{2\sigma_a^2}a_ia_i}\Bigg] \Bigg[ \frac{1}{V_d}\int_{V_d} db^d_{i}  \Bigg]   \Bigg[\frac{1}{2\pi} \int_{-\pi}^{\pi} dc_i \Bigg]e^{-\frac{\lambda}{4!}\int d^dx \phi^4(x)}}
\eea
with the implicit definition of the field  Eq.~\ref{eq:Act}.

%%%%%%%%%%%%%%%%%%%%%%%%%%%%%%%%%%%%%%%%%%%%%%%%%%%%%%%%%%%%%%%%%%%%%%%%%%%%%%%%%%%%%%%%%%%%
\subsection{Two point correlator in perturbation theory}
We now compute the two point correlator in perturbation theory to O($\lambda$) with correction up to O(1/N) terms. up to O$(\lambda)$ we can write our two point correlator as 
\bea
\langle \phi(w_1) \phi(w_2) \rangle &=&  \frac{\prod_{i=1}^N  \Bigg[ \int da_i  \frac{\sqrt{N}}{\sqrt{2\pi} \sigma_a} e^{-\frac{N}{2\sigma_a^2}a_ia_i}\Bigg]\Bigg[ \frac{1}{V_d}\int_{V_d} d^db_{i}  \Bigg] \Bigg[\frac{1}{2\pi} \int_{-\pi}^{\pi} dc_i \Bigg]\left(1-\frac{\lambda}{4!}\int d^dx \phi^4(x) \right) \phi(w_1) \phi(w_2) }{\prod_{i=1}^N  \Bigg[ \int da_i  \frac{\sqrt{N}}{\sqrt{2\pi} \sigma_a} e^{-\frac{N}{2\sigma_a^2}a_ia_i}\Bigg] \Bigg[ \frac{1}{V_d}\int_{V_d} d^db_{i}  \Bigg]   \Bigg[\frac{1}{2\pi} \int_{-\pi}^{\pi} dc_i \Bigg]\left(1-\frac{\lambda}{4!}\int d^dx \phi^4(x)\right)} \nn\\
& \equiv & \langle \phi(w_1) \phi(w_2)\rangle_{\text{f}} - \frac{\lambda}{4!}\int d^4 x\langle \phi(w_1) \phi(w_2) \phi^4(x)\rangle_{\text{f}} + \langle \phi(w_1) \phi(w_2)\rangle_{\text{f}} \frac{\lambda}{4!}\int d^4 x\langle  \phi^4(x)\rangle_{\text{f}} +O(\lambda^2) 
\label{eq:2pIT}
\eea
where the subscript `f' indicates correlators evaluated  in the free theory. 
 It is instructive to understand the various contributions to the 1/N corrections. 
We can identify two distinct types of contributions for computing corrections up to O(1/N). We see that the integral over $a_i$ are non zero for an integrand of $a_i^{\nu}$ weighted by the Gaussian weight $e^{-\frac{N}{2\sigma_a^2}a_ia_i}$ only when $\nu$ is even. It turns out that contribution to any correlation functions coming from integrands of the form $a_i^\nu e^{-\frac{N}{2\sigma_a^2}a_ia_i}$ with $\nu=0, 2$
captures all of $(1/N)^0$ piece as well as a part of the O$(1/N)$ piece.
The remaining $1/N$ contribution can be obtained by picking one of the Gaussian integral over $a_i$ to have $\nu=4$ with the rest over $a_{j\neq i}$ being $\nu=0, 2$.
\begin{itemize}
    \item  Integrals with at most $\nu=2$ naturally lead to diagrams in field theory that correspond to  pairwise contractions of the fields.   These give rise to a result proportional to both connected and disconnected diagrams from field theory. The proportionality factor is $N(N-1)..(N-j)/N^{j+1}$ for a correlator with (2(j+1)) fields. This can be seen in the explicit calculation for the two point correlator presented in Appendix \ref{App:twoP}. A consequence of this is that the disconnected diagrams which would usually be absent in connected correlators in quantum field theory, do not fully cancel between the contributions of the numerator and denominator of the connected correlator for the NNFT but will leave a correction at O(1/N) and higher.
    \item The other type of correction at O(1/N) comes from integrals  with 4 powers of $a_i$ (or $\nu=4$). These are non-Gaussian finite N corrections which have the same origin as the 4 point correlator calculation of the free theory NN. We will refer to these as four field / Non -Gaussian corrections.
\end{itemize}
Combining all the corrections, details can be found in Appendix \ref{App:twoP} we can write 
\bea 
\langle \phi(w_1) \phi(w_2) \rangle &=&\int_{V_d} \frac{d^db}{ (2\pi)^d}e^{i b \cdot(w_1-w_2)}\Bigg[ \frac{1}{b^2+m^2} -\frac{1}{(b^2+m^2)^2}\frac{\lambda}{2} \int \frac{d^dp}{(2\pi)^d}\left( \frac{1}{p^2+m^2}\right)\nn\\
&+& \frac{1}{(b^2+m^2)^2}\frac{3 \lambda}{2N} \int \frac{d^dp}{(2\pi)^d}\left( \frac{1}{p^2+m^2}\right) \nn\\
&+ & \frac{1}{b^2+m^2}\frac{1}{N}\frac{\lambda}{4}\delta^d(0)\left(\int \frac{d^dp}{(2\pi)^d} \frac{1}{p^2+m^2}\right)^2 \nn\\
&-&  \frac{1}{(b^2+m^2)^2}\left( \frac{3V_d}{8N} \left(2 \delta^d(0)  +\delta^d(2b) \right)\right)\int \frac{d^dp}{(2\pi)^d}\left( \frac{1}{p^2+m^2}\right)-\frac{3\lambda}{2N} \left(\frac{ V_d}{(2\pi)^d}\right)  \frac{1}{(b^2+m^2)^3} \Bigg]\nonumber\\
&\equiv & \int_{V_d} \frac{d^dp}{(2\pi)^d}e^{ip.(w_1-w_2)}G_2(p) 
\label{eq:2p}
\eea

The first line is the usual field theory result. The second line is the correction to the connected diagram, the third is the correction from non-cancellation of disconnected diagrams and the fourth line is contribution from the non-Gaussian/four field contraction.
For $w_i - w_j \sim 1/m$, then we can  estimate the scaling of each of these terms(for d $>$ 2) up to logarithmic corrections
\bea
&&\langle \phi(w_1) \phi(w_2) \rangle \sim m^{d-2}\Bigg[1- \frac{\lambda}{2} \frac{\Lambda^{d-2}}{m^2}+ \frac{3\lambda}{N}\frac{\Lambda^{d-2}}{m^2} +\frac{\lambda}{4N}\frac{V_s\Lambda^{2d-4}}{4}-\lambda\left(\frac{3}{4N}\Lambda^d V_s + \frac{3}{8N*2^dm^{d-4}} \right) \frac{\Lambda^{d-2}}{m^2}  -\frac{3\lambda}{2N}\frac{\Lambda^{d}}{m^4}\Bigg]\nn\\
\label{2p2}
\eea
Among these terms we see that the largest correction is of the form $\lambda\Lambda^d V_s/N \Lambda^{d-2}/m^2 $. 
 
After renormalization, since it is likely that some of these divergent terms can be absorbed into the bare parameters, namely $m, \lambda$ and the field strength renormalization factor Z. So we will come back and re-asses the error budget after implementing renormalization in Section \ref{sec:Renorm}.

We can likewise look at the scaling of the 1/N corrections in momentum space 
\bea
G_2(p) &=& \int d^dp e^{-i p \cdot( w_1-w_2)} \langle \phi(w_1) \phi(w_2) \rangle \sim \frac{1}{p^2+m^2}\left( 1- \frac{\lambda}{2}\frac{\Lambda^{d-2}}{p^2+m^2} \ln  \frac{\Lambda^2}{m^2}\right) \nn\\
&-&  \frac{\lambda}{N}\frac{1}{p^2+m^2}\left( \frac{\Lambda^{d-2}}{p^2+m^2} \ln  \frac{\Lambda^2}{m^2} \left( -3 +\frac{3}{2}V_dV_s+ \frac{3V_d}{4}\delta^d(2p)\right)+ \frac{3}{4}\frac{V_d}{(p^2+m^2)^2} -\frac{V_s\Lambda^{2d-4}}{4}\right)
\eea
If we are probing momenta of virtuality $p^2 \sim m^2$, then we see that the largest error comes from a term that scales as $V_dV_s \Lambda^{d-2}/(m^{d-2}N)$ .
We also see a term proportional to $\Lambda^{d-2}/(m^{d-2}N)V_d\delta^d(p)$ which  gives correction specifically at zero momentum. This is again following the same pattern as in free theory where the error adds up coherently at SKP in momentum space. 

%%%%%%%%%%%%%%%%%%%%%%%%%%%%%%%%%%%%%%%%%%%%%%%%%%%%%%%%%%%%%%%%%%%%%%%%%%%%%%%%%%%%%%
\subsection{Four point correlator}
We now look at the connected 4 point correlator defined as 
\bea
G_c^{(4)}(w_1, w_2, w_3, w_4) = \langle \phi(w_1) \phi(w_2) \phi(w_3) \phi(w_4) \rangle - \Big\{ \langle \phi(w_1) \phi(w_2) \rangle \langle \phi(w_3) \phi(w_4) \rangle + \text{permutations} \Big\}
\eea
up to order $1/N$. \\
{\bf Tree level:} 
We define the connected four point correlator at leading order in $\lambda$ and at leading order in $1/N$ as $G_c^{(4,1)}$ where the second entry on the superscript is denoting the order of the correction in $\lambda$. Thus, 
\bea
G_c^{(4,1)}(w_1, w_2, w_3, w_4)&=&-\lambda \int d^d x \prod_{i=1}^4\Bigg[ \int_{V_d} \frac{d^db_{i}}{(2\pi)^d}  \left(\frac{ \cos (b_{i} \cdot (w_i-x)) }{b_{i}^2+m^2}\right)\Bigg] \nn\\
&=& -\lambda \prod_{i=1}^4\Bigg[ \int_{V_d} \frac{d^db_{i}}{(2\pi)^d} \frac{ e^{b_i \cdot w_i}}{b_{i}^2+m^2}\Bigg]\delta^d(b_1+b_2+b_3+b_4)
\label{eq:tree}
\eea
which is just the field theory result. 
At next to leading order in $1/N$ at order $\lambda$ we again identify three distinct types of contributions 
\begin{enumerate}
    \item{{\bf Contribution proportional to connected field theory result}}
\bea
&&\frac{6\lambda}{N} \prod_{i=1}^4\Bigg[ \int_{V_d} \frac{d^db_{i}}{(2\pi)^d} \frac{ e^{b_i \cdot w_i}}{b_{i}^2+m^2}\Bigg]\delta^d(b_1+b_2+b_3+b_4)
\label{eq:treeN1}
\eea
\item{{\bf Contributions proportional to disconnected field theory diagrams}}

\begin{figure}
    \centering
    \includegraphics[width=0.5\linewidth]{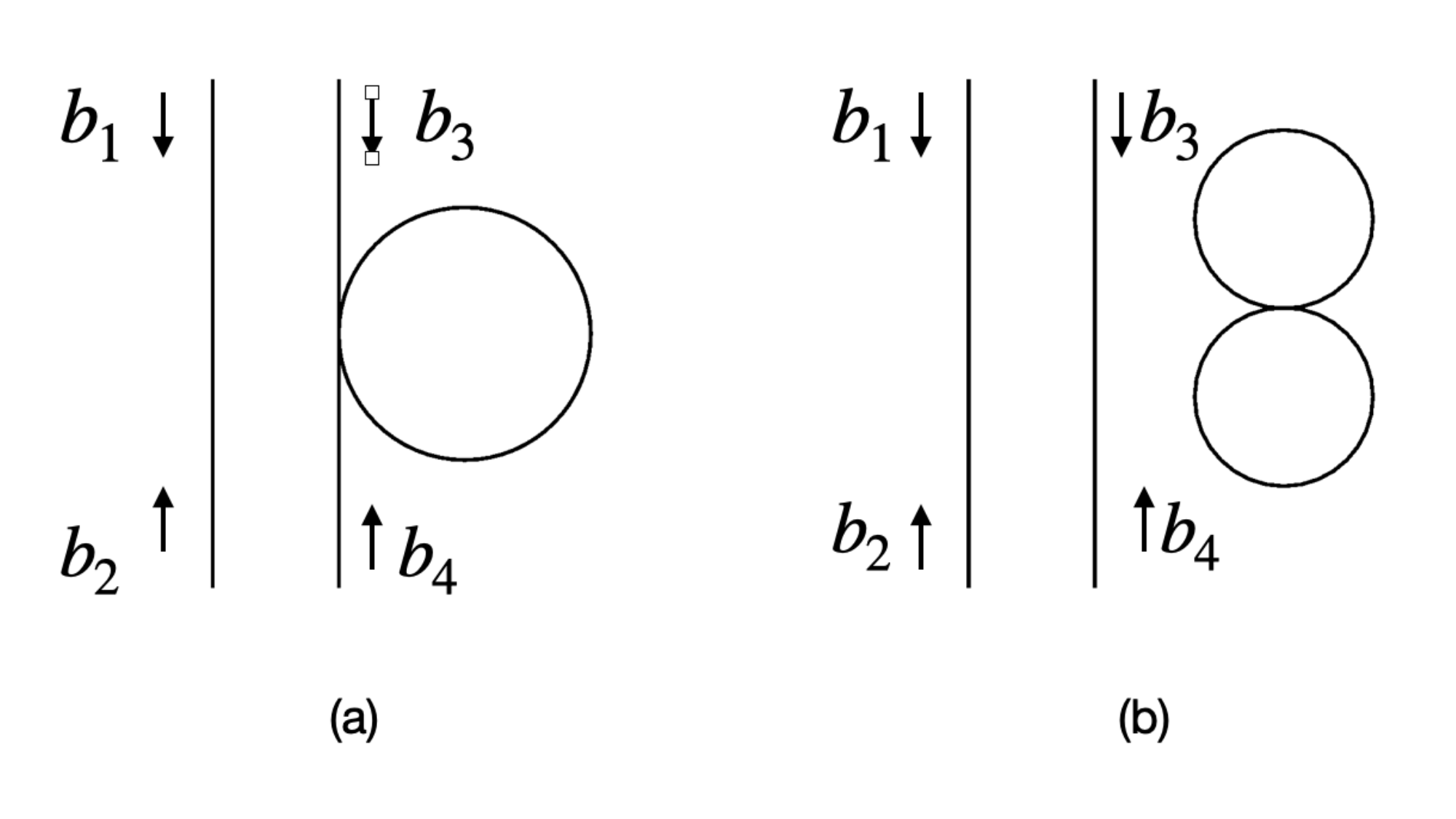}
    \caption{Partially and fully disconnected contributions to the 4 point correlator. The solid lines represent free field theory propagators.}
    \label{fig:dcNN}
\end{figure}
Just like the two point correlators, the contributions corresponding to disconnected field theory diagrams  at O(1/N) and higher powers of 1/N do not cancel out. The remainder will generally give contributions from both partially and fully disconnected diagrams. For e.g  for diagram shown in Fig.\ref{fig:dcNN}(a), we get the contribution 
\bea
\frac{3\lambda}{2N}\prod_{i=1}^4\Bigg[ \int_{V_d} \frac{d^db_{i}}{(2\pi)^d}  e^{ib_{i} \cdot w_i}\Bigg]\delta^d(b_{1}+b_{2}+b_{3}+b_{4})\frac{\delta^d(b_1-b_2)}{b_1^2+m^2}\frac{\delta^d(b_3-b_4)}{(b_3^2+m^2)^2}\int \frac{d^dp}{p^2+m^2}+  \text{permutations of $b_1, b_2,b_3,b_4$}\nn\\
\label{eq:treeN2}
\eea
We note that this gives a contribution only at SKPs, so that as long as we keep away from these points, these can be ignored. The same applies to diagram Fig.\ref{fig:dcNN}(b) for which however the 1/N correction completely cancels out. 

We will ignore these types of corrections from here on when we are interested in computing correlators in momentum space away from SKPs. Of course, these corrections are still relevant for position space correlators and in Section \ref{sec:IPT} we will explore NN architecture modifications to remove such contributions even at SKP. 

\item{{\bf Non-Gaussian/ four field contractions}}

There are several non zero contributions here and all of them can be combined together as 
\bea
&&-\frac{\lambda}{2N} \frac{V_d}{2(2\pi)^d}  \prod_{i=1}^4\Bigg[ \int_{V_d} \frac{d^db_{i}}{(2\pi)^d}  \left(\frac{ e^{ib_{i} \cdot w_i}}{b_{i}^2+m^2}\right)\Bigg]\delta^d(b_{1}+b_{2}+b_{3}+b_{4})\Bigg[\sum_{m \neq n }\Bigg[ \left( \delta^d(b_{m}-b_{n}) +2 \delta^d(b_{m}+b_{n}) \right)\Bigg]\nn\\
&+ &\sum_{m \neq n \neq q}(b_m^2+m^2)\Bigg[ \left( \delta^d(b_{m}-b_{n})\delta(b_{m}+b_{q}) + \delta^d(b_{m}+b_{n})\delta(b_{m}-b_{q}) + \delta^d(b_{m}+b_{n})\delta(b_{m}+b_{q}) \right)\Bigg]\int \frac{d^dp}{p^2+m^2}\Bigg]\nn\\
\label{eq:Tree4c}
\eea
We again see that these corrections only have support at SKPs. Hence when performing renormalization, we will ignore these terms. 
\end{enumerate}

The full result is then a sum of the terms in Eqs.~\ref{eq:tree}, \ref{eq:treeN1}, \ref{eq:treeN2}, \ref{eq:Tree4c}. 

We can take a look at the scaling of the 1/N corrections for position space correlators . If we are probing distances $w_i-w_j \sim 1/m$, then we estimate 
\bea
G_c(w_1,w_2, w_3, w_4) \sim \lambda m^{3d-8} \left(1+ \frac{6}{N} +\frac{3}{2}\frac{1}{N} \frac{\Lambda^{d-2}}{m^{d-2}}-\frac{9}{2N} \frac{\Lambda^d}{m^d}-\frac{3}{N} \frac{\Lambda^{2d-2}}{m^{2d-2}}\right) 
\eea
We see that just as for free theory, the largest error in position space scales as $\Lambda^{2d-2}/(m^{2d-2}N)$.  However as long as we keep away from the SKPs, we see that in momentum space, the only relevant correction is proportional to the field theory result and suppressed by a factor 1/N. \\

{\bf One loop:} 
In order to implement renormalization of the coupling, we will compute the $4$- point correlator at one-loop. Given our analysis of the O($\lambda$) result, we will only focus on corrections away from the SKPs. We will then renormalize and subsequently make predictions away from the SKPs. As before, we divide up the contributions into three categories, formally writing them as $C_{1,2,3}$ such that the full correlator is  

\bea 
G_{c}^{(4,2)}(w_1,w_2,w_3,w_4)=C_1+C_2+C_3
\eea 
\begin{enumerate}
    \item{{\bf Contribution proportional to the connected field theory result}}
    
     \begin{figure}
    \centering
    \includegraphics[width=0.6\linewidth]{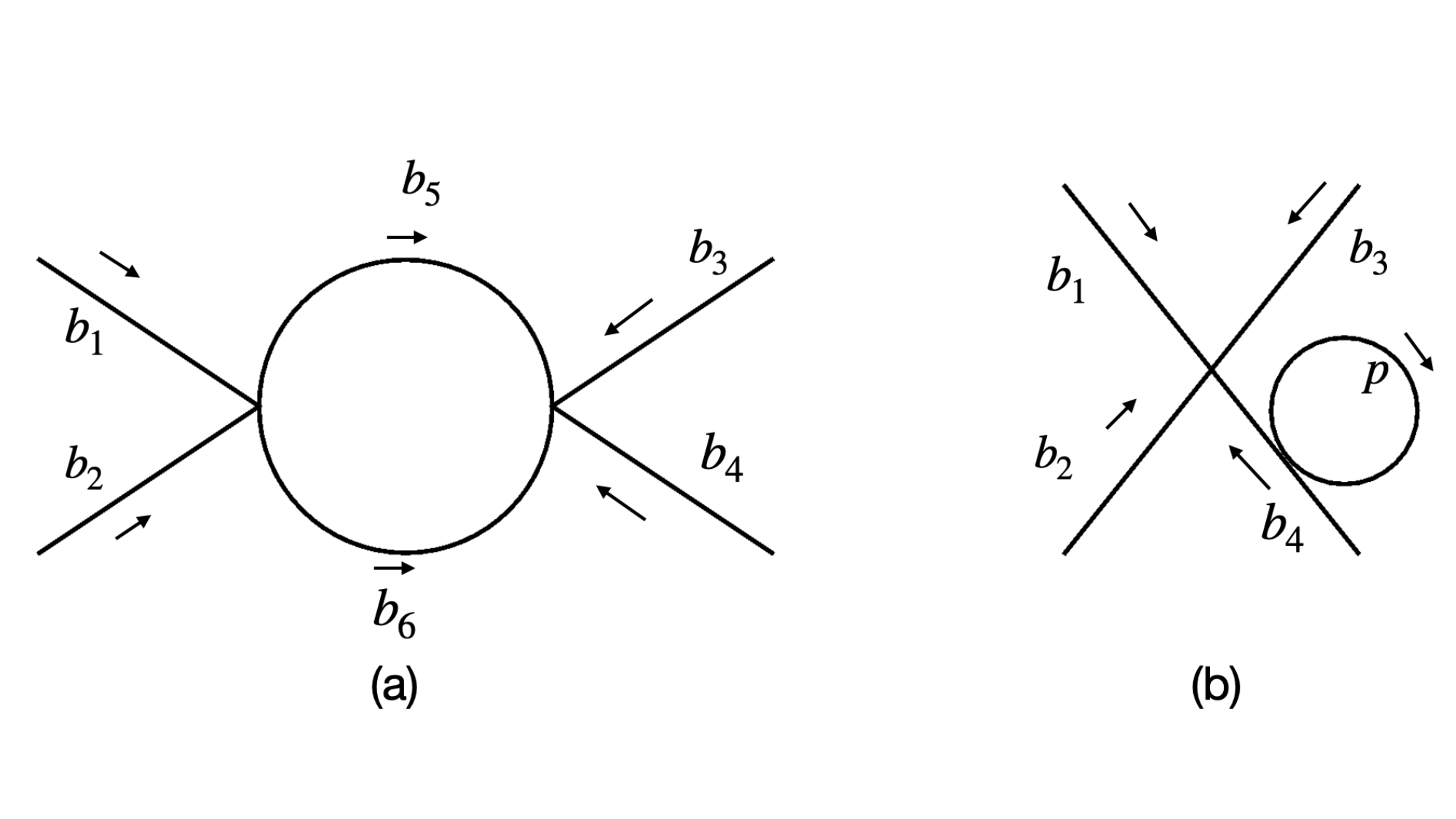}
    \caption{Connected diagrams from pairwise contractions of fields.}
    \label{fig:4pc}
\end{figure}
We denote the contributions proportional to the connected field theory result as $C_1$ which itself can be categorized in two types of terms, $C_{1,1}$ and $C_{1,2}$ where $C_{1,1}$ corresponds to the usual s, t and u channel connected diagrams shown in Fig.\ref{fig:4pc} (a). 
\bea 
C_{1,1}&=& \lambda^2\left(1-\frac{15}{N}\right) \prod_{i=1}^6\Bigg[\int_{V_d}\frac{ d^db_{i}}{(2\pi)^d}\frac{1}{b_i^2+m^2}\Bigg]\Bigg[\prod_{j=1}^4e^{w_j \cdot b_j}\Bigg]\delta^d(b_{1} +b_{2}-b_{5}-b_{6})\delta^d(b_{3}+b_{4}+b_{5}+b_{6})+ \text{u + t channel}\nonumber\\
\eea
This simplifies to

\bea
C_{1,1}&=& \lambda^2\left(1-\frac{15}{N}\right)\prod_{i=1}^4\Bigg[\int_{V_d}\frac{ d^db_{i}}{(2\pi)^d}\frac{e^{i w_i \cdot b_i}}{b_i^2+m^2}\Bigg]\delta^d(b_1+b_2+b_3+b_4)(V((b_1+b_2)^2 +V((b_1+b_3)^2) + V((b_1+b_4)^2)\nn\\
\label{eq:Loop}
\eea
where
\bea
V(b^2)=\frac{1}{2}\int_{V_d}\frac{ d^dp}{(2\pi)^d}\frac{1}{((b+p)^2+m^2)(p^2+m^2)}.
\label{eq:V}
\eea

Likewise we have contribution from external  legs (Fig.\ref{fig:4pc}(b)), which we denote as $C_{1,2}$ 
\bea
 C_{1,2}=\frac{\lambda^2}{2}\left(1- \frac{15}{N}\right)\prod_{i=1}^4\Bigg[ \int_{V_d} \frac{d^db_{i}}{(2\pi)^d} \frac{ e^{b_i \cdot w_i}}{b_{i}^2+m^2}\Bigg]\delta^d(b_1+b_2+b_3+b_4)\int_{V_d} \frac{d^dp}{(2\pi)^d}\frac{1}{p^2+m^2} \sum_{j=1}^4 \frac{1}{b_j^2+m^2}
 \label{eq:loopW}
\eea

\item{{\bf Contributions proportional to disconnected field theory diagrams}}
We denote contributions proportional to disconnected diagrams in quantum field theory as $C_2$.
    There are several diagrams of this type. But just as in the case of tree level, most of them will only contribute at SKPs and will have the momentum space form $\propto \delta^d(b_1-b_2)\delta^d(b_3-b_4)$. We therefore ignore them. There is however, one relevant contribution which survives away from SKPs at O$(1/N)$ corresponding to the diagram shown in Fig. \ref{fig:4pdc}. 
    \begin{figure}
    \centering
    \includegraphics[width=0.4\linewidth]{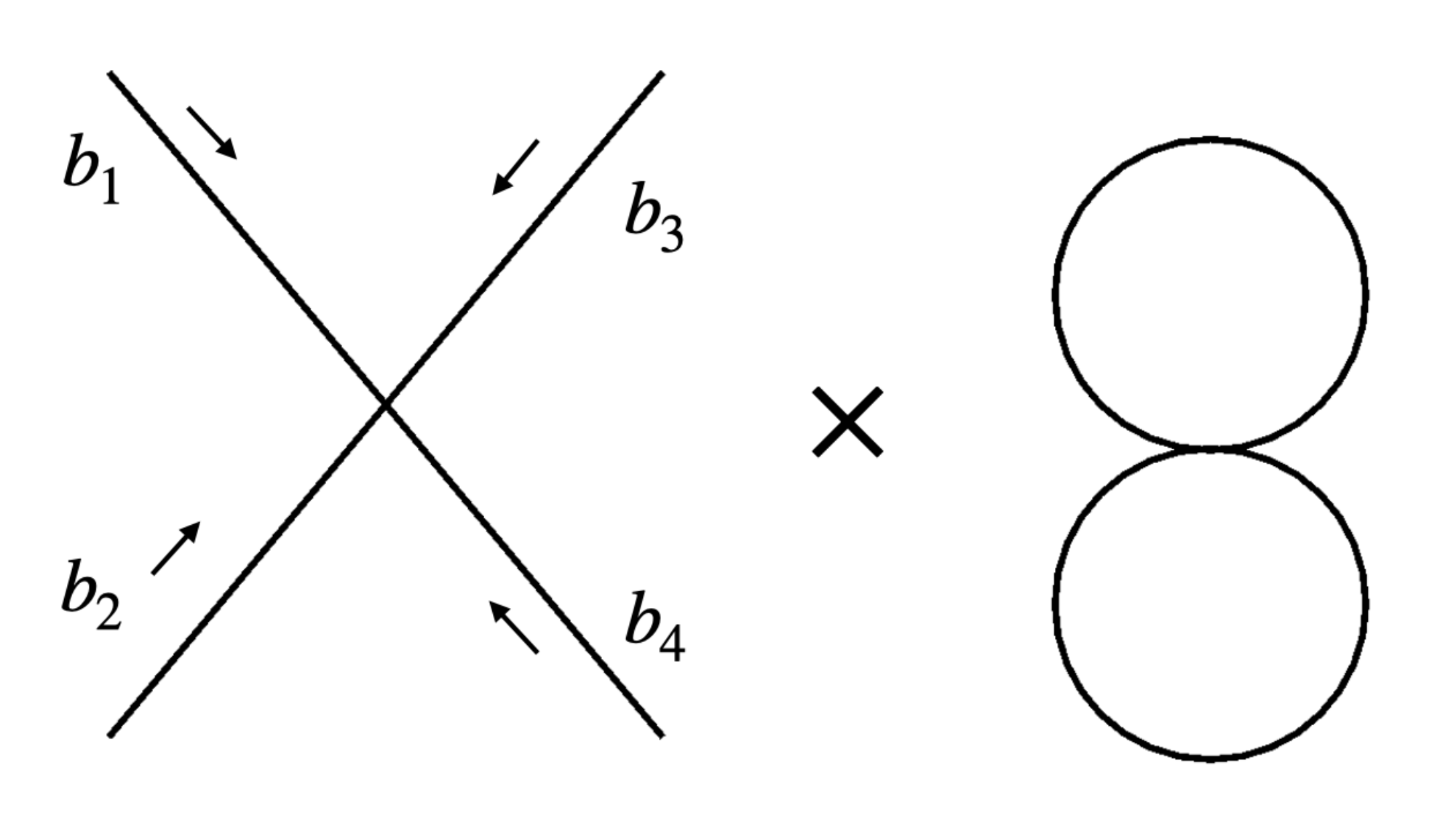}
    \caption{Disconnected diagram at O($\lambda^2$) that gives an O(1/N) correction to the connected 4 point correlator at generic external momenta.}
    \label{fig:4pdc}
\end{figure}
This gives us 
\bea 
C_2=-\frac{\lambda^2}{N}\prod_{i=1}^4\Bigg[ \int_{V_d} \frac{d^db_{i}}{(2\pi)^d} \frac{ e^{b_i \cdot w_i}}{b_{i}^2+m^2}\Bigg]\delta^d(b_1+b_2+b_3+b_4)\delta^d(0)\left(\int_{V_d}\frac{d^dp}{(2\pi)^d}\frac{1}{p^2+m^2}\right)^2
\label{eq:loopN1}
\eea

 \item{{\bf Contribution from four-field/non-Gaussian contractions}}
\\
As in the case of tree level result, we will obtain four-field/non-Gaussian contributions at SKPs as those in Eq.~\ref{eq:Tree4c}. Since we are away from SKPs none survive. There are however three other types of four field/non-Gaussian contribution that survive away from SKP. Together they give rise to our third contribution $C_3$. The three types of contributions have slightly different structure which is why decided to distinguish them as $C_{3,1}, C_{3,2}$ and $C_{3,3}$ such that 
\bea
C_3=C_{3,1}+C_{3,2}+C_{3,3}.
\eea

with 
\bea
    C_{3,1}&=&\lambda^2\frac{3}{4N}\prod_{i=1}^4\Bigg[ \int_{V_d} \frac{d^db_{i}}{(2\pi)^d} \frac{ e^{b_i \cdot w_i}}{b_{i}^2+m^2}\Bigg]\delta^d(b_1+b_2+b_3+b_4)\nonumber\\
    &&\left[\frac{V_d}{(b_1^2+m^2)}\Bigg[ \frac{2}{b_1^2+m^2} 
    + \frac{1}{2}\int \frac{d^dp}{(2\pi)^d}\frac{1}{p^2+m^2} \left( 2\delta^d(0)+ \delta^d(b_1)\right)\Bigg]+\{b_1\rightarrow b_2\}+\{b_1\rightarrow b_3\}+\{b_1\rightarrow b_4\}\right] \nonumber\\
    \label{eq:loopN2}
    \eea

    \bea
    C_{3,2}&=&\lambda^2\frac{3}{4N}\prod_{i=1}^4\Bigg[ \int_{V_d} \frac{d^db_{i}}{(2\pi)^d} \frac{ e^{b_i \cdot w_i}}{b_{i}^2+m^2}\Bigg]\delta^d(b_1+b_2+b_3+b_4)\sum_{b_i \neq b_j} \frac{V_d}{(b_i^2+m^2)}\Bigg[ \frac{2}{b_j^2+m^2}+ \frac{1}{(2b_i+b_j)^2+m^2}\Bigg]\nn\\
    \label{eq:loopN3}
    \eea
    where the sum $b_i,b_j$ runs over all 12 choices in all, any pair $b_i,b_j$ gives 6 choices and interchanging their order gives another 6.
    \bea
    C_{3,3}&=&\lambda^2\frac{3}{8N}\prod_{i=1}^4\Bigg[ \int_{V_d} \frac{d^db_{i}}{(2\pi)^d} \frac{ e^{b_i \cdot w_i}}{b_{i}^2+m^2}\Bigg]\delta^d(b_1+b_2+b_3+b_4)\Bigg[\frac{V_d}{(\left(\frac{b_1+b_2}{2}\right)^2+m^2)^2}+ (b_2 \rightarrow b_3) + (b_2 \rightarrow b_4) \Bigg]
    \label{eq:N4}
    \eea
We recognize some of the terms contributing to $C_{3,1}$ are the same type of contribution  to the two point correlator arising from non-Gaussianities.

\end{enumerate}

When external momenta are of the order of the mass scale $m$, the order of magnitude of the largest 1/N correction  compared to the tree level result for $\lambda \sim m^{d-4}$ scales as  
\bea
G_c^{(4,2)} \sim   \frac{V_d}{N m^{d}}.
\eea
However, this is still the bare result and the final error budget will be decided after renormalization which we discuss in the next section. 

%%%%%%%%%%%%%%%%%%%%%%%%%%%%%%%%%%%%%%%%%%%%%%%%%%%%%%%%%%%%%%%%%%%%%%%
\section{Perturbative renormalization}
\label{sec:Renorm}

We now look at the perturbative renormalization for $\phi^4$ theory on the neural network at finite N. We have two bare parameters m and $\lambda$ that need to be tuned to physical observables. 
Our objective is to determine the value of N such that 1/N corrections for observables after renormalization are small. 

{\bf Two-point correlator renormalization:}
We now have to choose a specific observable to tune the bare mass m. This can be for e.g., the correlation length which is the single particle pole of the two point correlator. 
Since we wish to keep away from kinematic points where the error is large, we choose instead the two point correlator at some specific non-zero momentum $\tilde p$ , $\chi_{2,\tilde p}$. To compare our bare perturbation theory result to the physical observable also requires us to implement field strength renormalization by introducing a Z factor defined as
\bea
G_{2,\text{ren}}(p) = Z^{-1} G_2(p)
\eea
where $G_2(p)$ is the Fourier transform of the two-point correlator given by 
\bea
G_2(p)=\int d^dx\langle \phi(x)\phi(0) \rangle e^{i p x}
\eea
and Z is extracted by choosing a suitable renormalization condition:
\bea
Z^{-1} = \frac{d}{dp^2}G^{-1}_2(p)\Big|_{p= \tilde p} \ \ \text{and}  \ \  G_{2,\text{ren}}(\tilde p) = \chi_{2,\tilde p}
\eea

At leading order in $\lambda $
    \bea 
  Z=1 \ \ \ \text{and} \ \ \chi_{2,\tilde p} = G_2(\tilde{p}) = \frac{1}{\tilde p^2+ m^2}  \implies m^2= \frac{1}{X_{2,\tilde p}}- \tilde p^2 \implies G_{2,\text{ren}}(p)= \frac{1}{p^2-\tilde p^2+ \frac{1}{X_{2,\tilde p}}}
    \eea
At next-to-leading order we obtain 
\bea 
Z^{-1}&=& 1- \frac{\lambda}{4N}\delta^d(0)\left(\int_{V_d} \frac{d^dq}{(2\pi)^d}\frac{1}{q^2+m^2}\right)^2- \frac{3\lambda}{2N}\frac{V_d}{(2\pi)^d} \frac{1}{(\tilde p^2+m^2)^2} \nn\\
\chi_{2,\tilde p} &=& \frac{1}{\tilde p^2+ m^2} -\frac{\lambda}{2}\left(1- \frac{3}{N} +\frac{3V_d}{2N}\delta^d(0)\right)\frac{1}{(\tilde p^2+m^2)^2 }\int \frac{d^dq}{(2\pi)^d}\frac{1}{q^2+m^2} -\frac{3\lambda}{N}\frac{V_d}{(2\pi)^d}\frac{1}{(\tilde p^2+m^2)^3}
\eea
which implies  a bare mass 
\bea
m^2= -\tilde p^2+ \frac{1}{\chi_{2,\tilde p}}- \frac{\lambda}{2}\left(1- \frac{3}{N} +\frac{3V_d}{2N}\delta^d(0)\right)\int \frac{d^dq}{(2\pi)^d}\frac{1}{q^2-\tilde p^2+ \frac{1}{\chi_{2,\tilde p}}}-\frac{3\lambda}{N}\frac{V_d}{(2\pi)^d}\chi_{2,\tilde p} +O(\lambda^2) 
\label{eq:bmass}
\eea
We can now rewrite our renormalized two point correlator ( ignoring the contribution at p=0)  as 

\bea
G_{2,\text{ren}}(p) = \frac{1}{p^2-\tilde p^2+ \frac{1}{X_{2,\tilde p}}}\Bigg[1 - \frac{3\lambda}{2N}\frac{V_d \chi_{2,\tilde p}^4}{(2\pi)^d}\frac{(p^2-\tilde p^2)^2}{(\chi_{2,\tilde p}(p^2-\tilde p^2)+ 1)^2}\Bigg] 
\eea
We then see that after renormalization the 1/N error  scales as $\lambda/N V_d \xi^4$, where $\xi \sim \chi_{2,\tilde p}$ is the correlation length implying the constraint $N \gg \Lambda^d \xi^d $ for the finite width errors to be small.

%%%%%%%%%%%%%%%%%%%%%%%%%%%%%%%%%%%%%%%%%%%%%%%%%%%%%%%%%%%%%%%%%%%%%%%%%%%%%%%%%%%%%%%%%%%%%%%%%%%%%%%%%%%%%%%%%%%%%%%%%%%%%%%%%%%%%%%%%%%%%%%%%%%%%%%
{\bf Four-point correlator renormalization:}
There is one more bare parameter to tune, the coupling strength $\lambda$ for which we can use the connected 4 point correlator $G_c^{(4)}$ as an observable. We have contributions from free theory (Eq.~\ref{eq:fourv2}), O($\lambda$) (Eqs.~\ref{eq:tree}, ~\ref{eq:treeN1},~\ref{eq:treeN2},~\ref{eq:Tree4c}) and at O($\lambda^2$)( Eqs.~\ref{eq:Loop}, ~\ref{eq:loopW},~\ref{eq:loopN1},~\ref{eq:loopN2},~\ref{eq:loopN3},~\ref{eq:N4}). 
We have access to the $4-$ point correlator in the NN data. We can use this data to extract the four-point amplitude which we are going to renormalize away from the SKPs such that $|\tilde p_1|\neq |\tilde p_2|\neq |\tilde p_3| \neq |\tilde p_4|$. This automatically excludes several of the most singular contributions. 
To proceed to four-point correlator renormalization we begin with the definition
\bea
G^{(4)}_{c,\text{ren}}= Z^{-2}G_{c}^{(4)}
\eea
and define our observable $\chi_{4,\tilde p}$
\bea
\chi_{4,\tilde p}&=& \prod_{i=1}^4(\tilde p_i^2-\tilde p^2+\chi_{2,\tilde p}^{-1}) \frac{G_{c,\text{ren}}^{(4)} (\tilde p_1,\tilde p_2,\tilde p_3,\tilde p_4) }{\delta^d(\tilde p_1+\tilde p_2+\tilde p_3+\tilde p_4)}=Z^{-2}\prod_{i=1}^4(\tilde p_i^2-\tilde p^2+\chi_{2,\tilde p}^{-1}) \frac{G_{c}^{(4)} (\tilde p_1,\tilde p_2,\tilde p_3,\tilde p_4) }{\delta^d(\tilde p_1+\tilde p_2+\tilde p_3+\tilde p_4)} \nn\\
&= & \prod_{i=1}^4 \frac{(\tilde p_i^2-\tilde p^2+\chi_{2,\tilde p}^{-1})}{\tilde p_i^2+m^2}\Bigg[ -\lambda\left(1-\frac{6}{N}\right) +\lambda^2\left(1-\frac{15}{N}\right)\Big[V({\tilde  s})+V({\tilde  t})+V({\tilde  u})\Big]\nn\\
&+& \frac{\lambda^2}{2} \left(1-\frac{15}{N}\right) \int_{V_d}\frac{d^dq}{(2\pi)^d}\frac{1}{q^2+m^2}\sum_j \frac{1}{\tilde p_j^2+m^2} -\frac{\lambda^2}{2N}\delta^d(0)\left(\int_{V_d}\frac{d^d q}{(2\pi)^d}\frac{1}{q^2+m^2}\right)^2 \nn\\
&+& \frac{3\lambda^2}{4N}V_d \sum_j\left( \frac{1}{\tilde p_j^2+m^2} \left(\frac{2}{\tilde p_j^2+m^2} +\delta^d(0) \int \frac{d^dq}{(2\pi)^d} \frac{1}{q^2+m^2} \right)\right) \nn\\
&+& \frac{3\lambda^2}{4N}\sum_{j\neq k}\Big[\frac{V_d}{\tilde p_j^2+m^2}\left(\frac{2}{\tilde p_k^2+m^2}+\frac{1}{(2\tilde p_j+\tilde p_k)^2+m^2}\right)+\frac{1}{8}\frac{V_d}{\frac{(\tilde p_j+\tilde p_k)^2}{4}+m^2}\Big]+ \frac{3\lambda^2}{N}\frac{V_d}{(2\pi)^d}\frac{1}{(\tilde p^2+m^2)^2}\Bigg]
\eea
where in the last line the sum $j \neq k$ runs over 12 choices of ordered $b_i, b_j$ pairs.
To simplify this further, we need to plug in the definition of the bare mass, and expand the result to $O(\lambda^2)$ which gives us 
\bea
\chi_{4,\tilde p}&= &  -\lambda\left(1-\frac{6}{N}\right) +\lambda^2\left(1-\frac{15}{N}\right)\Big[V({\tilde s})+V({\tilde t})+V({\tilde u})\Big] \nn\\
&-& \frac{3\lambda^2}{N}  \int_{V_d}\frac{d^dq}{(2\pi)^d}\frac{1}{q^2-\tilde p^2+\chi_{2,\tilde p}^{-1}}\sum_j \frac{1}{\tilde p_j^2-\tilde p^2+\chi_{2,\tilde p}^{-1}} -\frac{\lambda^2}{2N}\delta^d(0)\left(\int_{V_d}\frac{d^d q}{(2\pi)^d}\frac{1}{q^2-\tilde p^2+\chi_{2,\tilde p}^{-1}}\right)^2 \nn\\
&+& \frac{3\lambda^2}{2N}V_d \sum_j\left( \frac{1}{\tilde p_j^2-\tilde p^2+\chi_{2,\tilde p}^{-1}} \left(\frac{1}{\tilde p_j^2-\tilde p^2+\chi_{2,\tilde p}^{-1}} -2\chi_{2,\tilde p}\right)\right) \nn\\
&+& \frac{3\lambda^2}{4N}\sum_{j\neq k}\Big[\frac{V_d}{\tilde p_j^2-\tilde p^2+\chi_{2,\tilde p}^{-1}}\left(\frac{2}{\tilde p_k^2-\tilde p^2+\chi_{2,\tilde p}^{-1}}+\frac{1}{(2\tilde p_j+\tilde p_k)^2-\tilde p^2+\chi_{2,\tilde p}^{-1}}\right)+\frac{1}{8}\frac{V_d}{\frac{(\tilde p_j+\tilde p_k)^2}{4}-\tilde p^2+\chi_{2,\tilde p}^{-1}}\Big]\nn\\
&+& \frac{3\lambda^2}{N}\frac{V_d}{(2\pi)^d}\chi_{2,\tilde p}^2
\label{eq:chi4p}
\eea
We see something interesting here that even after tuning the bare mass, we still have a O(1/N) 1PI correction that remains uncanceled. This is suggesting that even after renormalization, there are divergences which remain when computing other observables. We will discuss this in more detail in Section \ref{sec:HighPT} when we discuss higher point correlators.
We can now solve for the bare coupling $\lambda$ in terms of our observable
\bea
\lambda &=& -\chi_{4,\tilde p} \left(1+\frac{6}{N}\right) + \chi_{4,\tilde p}^2\Bigg[ \left(1+\frac{3}{N}\right)\Big[V({\tilde s})+V({\tilde t})+V({\tilde u})\Big] \nn\\
&-& \frac{3}{N}  \int_{V_d}\frac{d^dq}{(2\pi)^d}\frac{1}{q^2-\tilde p^2+\chi_{2,\tilde p}^{-1}}\sum_j \frac{1}{\tilde p_j^2-\tilde p^2+\chi_{2,\tilde p}^{-1}} -\frac{\lambda^2}{2N}\delta^d(0)\left(\int_{V_d}\frac{d^d q}{(2\pi)^d}\frac{1}{q^2-\tilde p^2+\chi_{2,\tilde p}^{-1}}\right)^2 \nn\\
&+& \frac{3}{2N}V_d \sum_j\left( \frac{1}{\tilde p_j^2-\tilde p^2+\chi_{2,\tilde p}^{-1}} \left(\frac{1}{\tilde p_j^2-\tilde p^2+\chi_{2,\tilde p}^{-1}} -2\chi_{2,\tilde p}\right)\right) \nn\\
&+& \frac{3}{4N}\sum_{j\neq k}\Big[\frac{V_d}{\tilde p_j^2-\tilde p^2+\chi_{2,\tilde p}^{-1}}\left(\frac{2}{\tilde p_k^2-\tilde p^2+\chi_{2,\tilde p}^{-1}}+\frac{1}{(2\tilde p_j+\tilde p_k)^2-\tilde p^2+\chi_{2,\tilde p}^{-1}}\right)+\frac{1}{8}\frac{V_d}{\frac{(\tilde p_j+\tilde p_k)^2}{4}-\tilde p^2+\chi_{2,\tilde p}^{-1}}\Big]\nn\\
&+& \frac{3}{N}\frac{V_d}{(2\pi)^d}\chi_{2,\tilde p}^2\Bigg]
\eea
Finally we can give our prediction for the 4 point correlator (again avoiding the SKPs ) 
\bea
 \chi_{4,k}&=& \prod_{i=1}^4(k_i^2-\tilde p^2+\chi_{2,\tilde p}^{-1}) \frac{G_{c,\text{ren}}^{(4)} (k_1,k_2,k_3,k_4) }{\delta^d(k_1+k_2+k_3+k_4)}\nn\\
 &= &  \chi_{4,\tilde p} + \chi_{4,\tilde p}^2 \Big[ Q(k_1,k_2,k_3,k_4)- Q(\tilde p_1, \tilde p_2, \tilde p_3,\tilde p_4 \Big]
 \label{eq:4pRen}
\eea
where 
\bea
&&Q(k_1,k_2,k_3,k_4) = \left(1-\frac{3}{N}\right) (V(s)+ V(u)+V(t)) - \frac{3}{N}  \int_{V_d}\frac{d^dq}{(2\pi)^d}\frac{1}{q^2-\tilde p^2+\chi_{2,\tilde p}^{-1}}\sum_j \frac{1}{k_j^2-\tilde p^2+\chi_{2,\tilde p}^{-1}}  \nn\\
&+& \frac{3}{2N}V_d \sum_j\left( \frac{1}{k_j^2-\tilde p^2+\chi_{2,\tilde p}^{-1}} \left(\frac{1}{k_j^2-\tilde p^2+\chi_{2,\tilde p}^{-1}} -2\chi_{2,\tilde p}\right)\right) \nn\\
&+& \frac{3}{4N}\sum_{j\neq l}\Big[\frac{V_d}{k_j^2-\tilde p^2+\chi_{2,\tilde p}^{-1}}\left(\frac{2}{k_l^2-\tilde p^2+\chi_{2,\tilde p}^{-1}}+\frac{1}{(2k_j+k_l)^2-\tilde p^2+\chi_{2,\tilde p}^{-1}}\right)+\frac{1}{8}\frac{V_d}{\frac{(k_j+k_l)^2}{4}-\tilde p^2+\chi_{2,\tilde p}^{-1}}\Big]\nn\\
\label{eq:4pQ}
\eea
We can then verify that the largest finite N error compared to the field theory result scales as $\Lambda^d \xi^d/N$ just as for the two point correlator.

%%%%%%%%%%%%%%%%%%%%%%%%%%%%%%%%%%%%%%%%%%%%%%%%%%%%%%%%%%%%%%%%%%%%%%%%%%%%%%%%%%%%%%%%%%%%%%%%%%%%%%%%%%%%%%%%%%%%%%
\section{Improved perturbation theory}
 \label{sec:IPT}

As seen in the previous sections, for our final prediction after tuning all the bare parameters we have three types of finite N corrections left 
\begin{enumerate}
    \item The first term in the expression for $Q(k_1,k_2,k_3,k_4)$ in Eq. \ref{eq:4pQ} is a contribution proportional to the field theory prediction. This is what we expect; the finite N corrections are suppressed by O(1/N) compared to the true result. 
    \item The second term in Eq. \ref{eq:4pQ} is disturbing since it is a remnant of the one loop (1PI) correction to the  two point correlator of external legs (Fig.\ref{fig:4pc}(b)); normally one would expect that for any observable in a renormalizable theory these types of 1PI divergences would be fully absorbed into the bare mass. This signals to us the non-renormalizability of this theory. We see that this correction scales as $\Lambda^{d-2}\xi^{d-2}/N$ compared to the exact field theory result. 
    Hence even after renormalization, the four-point correlator is sensitive to the UV cut-off.  This non cancellation therefore prompts us to conjecture that at any higher order in perturbation theory, we may expect an O( $1/N(\Lambda^{d-2}\xi^{d-2})^n$)  correction coming from incomplete cancellation of the n-PI diagram. This needs to be confirmed by explicit calculation and we leave this for the future. This would then suggest that the perturbative expansion parameter for the 1/N correction is actually $\lambda \Lambda^{d-2}\xi^{d-2} $ which would restrict the validity of this expansion to $ \chi_{4p} \ll 1/(\Lambda^{d-2}\xi^{d-2})$. Further to maintain accuracy to O(n) in perturbation theory we would need $N \gg (\Lambda^{d-2}\xi^{d-2})^n$.
    \item Finally we have corrections that scale as $\Lambda^d\xi^d/N$ from the non-Gaussian correlations induced by the finite width NN. This scaling holds at any order in perturbation theory which suggest that as long as $N \gg \Lambda^d\xi^d$, these corrections are small and under control. 
\end{enumerate}

We conclude that the uncanceled $1/N$ suppressed UV divergent 1PI diagrams are main obstacle to the viability of scalar perturbation theory. The obvious solution is to modify the theory on the NN side to remove these divergent $\Lambda $ dependent contributions. At one loop the only contribution to 1PI are the bubble diagrams and we now modify our NN architecture to remove these bubble diagrams to all orders in perturbation theory. In the exact field theory result, bubble contributions are always absorbed into the bare mass. The modification on the NN side involving bubble removal will therefore change the equation for the bare mass while leaving the renormalized result for the exact field theory unchanged. 
We can further improve the convergence of the NN field theory by removing the non-gaussian corrections. Removal of the bubble diagram as well as non-Gaussian contributions can be implemented by modifying the NN probability distribution function through 
\bea
&& \frac{\lambda}{4!}\int d^dx \phi^4(x) \rightarrow \frac{\lambda}{4!}\int d^dx \Bigg[\phi^4(x)- 6\sum_{k}^N \tilde \phi^2_{k}(x)\sum_{k_1\neq k, k_2}^N \tilde \phi_{k_1}(x) \sum_{k_2 \neq k,k_1}^N \tilde \phi_{k_2}(x) -4 \sum_{k}^N \tilde \phi^3_k(x) \sum_{k_1\neq k}^N \tilde \phi_{k_1}(x)\nn\\
&- & 3 \sum_{k_1}^N \tilde \phi^2_{k_1}(x)\sum_{k_2\neq k_1}^N \tilde \phi^2_{k_2}(x)- \sum_{k}^N \tilde \phi^4_{k}(x) \Bigg]
\label{eq:Pmod}
\eea
where 
\bea 
\tilde \phi_{k}(x)= \frac{\sqrt{2V_d}}{\sigma_a(2\pi)^{d/2}}\frac{a_{k}\cos(b_{k}\cdot x +c_{k})}{\sqrt{b_{k}^2+m^2}}.
\eea
In Eq.~\ref{eq:Pmod} the first term in the modification removes the bubble contributions, the second third  and fourth terms remove non-gaussian corrections while the third term also removes disconnected diagrams. We recognize that the expression in Eq. \ref{eq:Pmod} can be equivalently written in a compact manner as 
\bea 
\frac{\lambda}{4!}\int d^dx \phi^4(x) \rightarrow \frac{\lambda}{4!}\int d^dx \prod_{i=1}^4 \sum_{k_i=1}^N\tilde \phi_{k_i}(x)\Big|_{k_i \neq k_j }
\label{eq:ModP}.
\eea
i.e. we only keep the contributions where the fields $\phi_k(x)$ are effectively evaluated on distinct neurons.

%%%%%%%%%%%%%%%%%%%%%%%%%%%%%%%%%%%%%%%%%%%%%%%%%%%%%%%%%%%%%%%%%%%%%%%%%%%%%%%%%%%%%%%%%%%%%%%%%%%%%%%%%%%%%%%%%%

With this modification, the two point correlator to O($\lambda$) is 
\bea 
&&\langle \phi(w_1) \phi(w_2) \rangle =\int_{V_d} \frac{d^db}{ (2\pi)^d}e^{i b \cdot(w_1-w_2)} \frac{1}{b^2+m^2}. 
\label{eq:2pM}
\eea
Note that along with the 1/N corrections, we have also removed all contributions from the field theory bubble diagram. The renormalization is therefore trivial and we have 
\bea
m^2 =- \tilde p^2+ \frac{1}{\chi_{2,\tilde p }},  \ \ \ Z=1.  
\label{eq:Impm}
\eea
The bare connected four point correlator becomes 
\bea
\chi_{4,p}&= &  -\lambda\left(1-\frac{6}{N}\right) +\lambda^2\left(1-\frac{15}{N}\right)\Big[V({s})+V({t})+V({ u})\Big]. 
\eea
After renormalization, we have 
\bea
\lambda &=& -\chi_{4,\tilde p} \left(1+\frac{6}{N}\right) + \chi_{4,\tilde p}^2\Bigg[ \left(1+\frac{3}{N}\right)\Big[V({\tilde s})+V({\tilde t})+V({\tilde u})\Big] 
\label{eq:Impc}
\eea
which then gives us our prediction for the observable $\chi_{4,k}$ at momenta different from $\tilde{p}_i$
\bea
 \chi_{4,k}&=& \prod_{i=1}^4(k_i^2-\tilde p^2+\chi_{2,\tilde p}^{-1}) \frac{G_{c,\text{ren}}^{(4)} (k_1,k_2,k_3,k_4) }{\delta^d(k_1+k_2+k_3+k_4)}\nn\\
 &= &  \chi_{4,\tilde p } + \chi_{4,\tilde p}^2 \Big[ Q(k_1,k_2,k_3,k_4)- Q(\tilde p_1, \tilde p_2, \tilde p_3,\tilde p_4 \Big]
\eea
where 
\bea
&&Q(k_1,k_2,k_3,k_4) = \left(1-\frac{3}{N}\right) (V(s)+ V(u)+V(t)) 
\eea
where the finite width corrections are restricted to those proportional to the field theory result and suppressed by $N$.

At this stage, it may be useful to consider how the cost of sampling changes as one modifies the distribution function according to Eq. \ref{eq:ModP}. For simplicity, we analyze the chose of sampling the parameter $b_i$ as we change the distribution. Similar analysis hold for the parameters $a_i$ and $c_i$. 
We can consider a one dimensional example for illustration. To begin with let's consider the cost of sampling for the standard form of the $\phi^4$ interaction as shown in the LHS of Eq. \ref{eq:ModP}.  Here we sample $N$ parameters $\{b_i\}_{i=1}^N$ from a joint probability distribution of the form :
\begin{equation}
p(\mathbf{b}) \propto \exp\Bigg[
-\sum_{i=1}^N \frac{b_i^2}{C}
+ \int dx \,\,\phi^4(x)
\Bigg]
\end{equation}
where $\phi(x)$ is sum over a specific function of the $b_i's$ as in Eq. \ref{eq:Act}. If the integral is discretized with $N_x$ points, then the computational cost scale as O($N*N_x$). In the improved architecture which removes all bubble diagrams, the probability distribution gets modified to 
\begin{equation}
p(\mathbf{b}) \propto \exp\Bigg[
-\sum_{i=1}^N \frac{b_i^2}{C}
+ \int dx \prod_{i=1}^4\sum_{k_i=1}^N \tilde \phi_{k_i}(x) \Big|_{k_1 \neq k_2 \neq k_3\neq k_4}
\Bigg]
\end{equation}
This defines a highly coupled distribution with apparent four-body interactions whose naive computational cost would be O($N^4*N_x$). However, we can restructure this by first defining power sums:
\[
S_1 = \sum_i \tilde \phi_{k_i},\quad
S_2 = \sum_i \tilde \phi^2_{k_i},\quad
S_3 = \sum_i \tilde \phi^3_{k_i},\quad
S_4 = \sum_i \tilde \phi^4_{k_i}
\]
Then the distinct-index sum can be rewritten as:
\[
\sum_{\text{distinct } i,j,k,l} \tilde \phi_{k_i} \tilde \phi_{k_j} \tilde \phi_{k_k} \tilde \phi_{k_l}
=
S_1^4
- 6 S_1^2 S_2
+ 3 S_2^2
+ 8 S_1 S_3
- 6 S_4
\]
For each term $S_1,S_2,S_3, S_4$  computational cost is O($N$), hence the cost per computation still scales as O($N*N_x$).

%%%%%%%%%%%%%%%%%%%%%%%%%%%%%%%%%%%%%%%%%%%%%%%%%%%%%%%%%%%%%%%%%%%%%%%%%%%%%%%%%%%%%%%%%%%%%%%%%%%%%%%%%%%%%%%%%%%%%%%%%%%%%%%%%%%%%%%%%%
 \section{Higher order analysis}
 \label{sec:HighPT}

\subsection{Errors at higher orders in $\lambda$}
So far our analysis has been limited to two and four point correlators to one loop. We see that with appropriate modification of the probability distribution, and by working away from SKPs, we can keep finite width errors under control at least to one loop. In this section, we want to discuss, to what extent this applies to arbitrary order in perturbation theory. We will also discuss errors for higher point correlators. 

From the tree level and one loop analysis we see that the leading order $1/N$ errors come from three sources. 
\begin{enumerate}
    \item{{\bf Corrections proportional to connected field theory diagrams}:}\\
    These corrections appear strictly from pairwise groupings of $a_{k_i}$ in the integrand of the NN integral, also referred to as two field contractions. This effectively translates to each pair $a_{k_i}$s being evaluated on a distinct neuron. This naturally leads to field theory propagators coming from Wick contractions. Since we have a finite number of neurons, for a contraction of $2n$ fields, the requirement for each pair to be evaluated on a distinct neuron leads to a combinatoric factor $N(N-1)..(N-n+1)$ and a corresponding result proportional to $N(N-1)..(N-n+1)/N^n$.  Perturbation theory  is usually an asymptotic expansion valid till O($1/\lambda$) so that if we are working to some fixed order $n \leq 1/\lambda$, we need $ N \gg n$ to keep the errors under control. 
    
    This also means that corrections proportional to powers of the 1PI diagrams \textit{do not} form a simple geometric series. Hence these corrections, which for instance in 4D $\phi^4$ theory contain powers of $\lambda \Lambda^2\xi^2$ cannot be fully absorbed in the bare mass. This can be easily seen as follows. 
    Consider the two point correlator at O($\lambda^2$). 
   One of the 1/N corrections  is a term proportional to the square of the $O(\lambda)$ 1PI diagram. This will scale as $N(N-1)(N-2)(N-3)(N-4)/N^5 \times (O(\lambda)_{1PI})^2$ leading to a correction $( 1 -10/N )(O(\lambda)_{1PI})^2 +O(1/N^2) $. For a renormalizable theory we expect this to be absorbed in the bare mass. However using our solution for the bare mass Eq.~\ref{eq:bmass}, we see that expanding out the renormalized two point correlator to O($\lambda^2)$ yields a contribution $(1- 6/N)(O(\lambda)_{1PI})^2$ which clearly does not capture the 1/N correction. This means that the two point correlator will suffer from corrections $1/N ( \lambda \Lambda^{d-2}\xi^{d-2})^n$ as we go to higher orders in perturbation theory. For perturbative expansion to be valid, we need $\lambda \Lambda^{d-2} \xi^{d-2}\ll 1 $. Likewise for accurately recovering the field theory result to O($\lambda^n$), we would need $N \gg (\Lambda^2 \xi^2)^n $.  
   Implementing improved perturbation theory pushes these corrections to one order higher, since we only start from 1PI diagrams at O($\lambda^2$) leading to the  constraint $\lambda \Lambda \xi\ll 1 $  in $d=4$ and $\lambda \ln \Lambda \xi\ll 1 $ in $d=3$ resulting in improved convergence of perturbation theory.

    \item{{\bf Corrections proportional to disconnected field theory diagrams}:} \\
    We also see from explicit calculations that the disconnected field theory diagrams do not cancel at $O(1/N)$ or higher. We see that such corrections proportional to $V_s$ can only appear through an x independent  $\int d^d x$ integral from the insertion of the interaction term. Thus with a modified probability distribution (Eq.~\ref{eq:ModP}), we can eliminate disconnected contributions proportional to the space-time volume $V_s$ to all orders in perturbation theory. The remaining disconnected pieces, for e.g. the partially disconnected diagrams (for at least the two and four point correlators) like Fig.~\ref{fig:dcNN}(a)  will always be at the SKPs and by working away from such points we completely eliminate any contribution to the error from such diagrams to all orders in perturbation theory. 
    \item{{\bf Non-Gaussian corrections from contractions of four or more fields}:}\\
    The third type of error appears from contractions of $4$ or more fields ($a_{k_i}$s). This corresponds to contribution from $4$ or more of the fields ($a_{k_i}$s) effectively being evaluated on the same neuron. Contraction of exactly four fields ($a_{k_i}$s) yields a $1/N$ error, contraction of $6$ fields ($a_{k_i}$s) or two separate groups of $4$ fields ($a_{k_i}$s) will lead to a $1/N^2$ error and so on.  The modification proposed in Eq.~\ref{eq:Pmod} ensures the cancellation of all such correction to $O(\lambda^2)$. While this will remove a large portion of these errors at higher order in perturbation theory as well, for now there does not appear to be any simple modification to the probability distribution that will remove \textit{all} such errors. For example, at O($\lambda^4$) we can have contractions of 4 fields, drawing one from each insertion of the interaction term. Therefore, we conjecture that at higher orders we may still have uncanceled non-Gaussian corrections that will lead to errors that scale as $(\Lambda ^d \xi^d/N)^n$. Hence to maintain accuracy we would need $N \gg \Lambda ^d \xi^d$. This will also be true for the four point correlator. 
    
\end{enumerate}
%%%%%%%%%%%%%%%%%%%%%%%%%%%%%%%%%%%%%%%%%%%%%%%%%%%%%%%%%%%%%%%%%%%%%%%%%%%%%%%%%%%%%%%%%%%%%%%%%%%%%%%
\subsection{Higher point correlators}    
When the dust settles, at each order in perturbation theory, the O(1/N) errors for the renormalized result of the two and 4 point correlators consist of three types of contributions: 1. Proportional to the renormalized field theory 2. Powers of 1PI diagrams which will be sensitive to the UV cut-off. 3. Non-Gaussian  corrections  proportional to powers of $\lambda \Lambda^d\xi^d/N$.

The next question we want to address whether this conclusion holds for higher point correlators. 
We will show that apart from the three types of corrections discussed in the previous paragraph, the higher point correlators will have additional divergences proportional to the \textit{bare} field theory result suppressed by N which will not be absorbed by the tuning of the bare parameters that was implemented using the two and four point correlators. 
To illustrate this, we consider the computation of the 6 point correlator in the improved perturbation theory framework. At tree level, ignoring corrections from SKPs we get 
\bea
G_{c}^{(6)}(k_1,...,k_6)&=& \lambda^2\left( 1- \frac{21}{N} \right)\frac{1}{(k_1+k_2+k_3)^2+m^2}\nn\\
&=& \chi_{4,\tilde p}^2\left(1-\frac{9}{N}\right)\frac{1}{(k_1+k_2+k_3)^2-\tilde p^2+\frac{1}{\chi_{2,\tilde p}}} + \text{permutations}
\eea
A full one loop calculation is beyond the scope of this paper, but to illustrate our point it is sufficient to keep track of corrections proportional to the field theory result while ignoring those from non-gaussian 4 field contractions (which scale as $\Lambda^d \xi^d/N$) 
\bea
&&G_{c}^{(6)}(k_1,...,k_6)= 
\lambda^2\left(1-\frac{21}{N}\right)\frac{1}{(k_1+k_2+k_3)^2+m^2}  \nn\\
&-& \lambda^3 \left(1 - \frac{36}{N}\right) \frac{\Bigg[V(s_{12})+V(s_{13})+V(s_{23})+V(s_{45})+V(s_{46})+V(s_{56})\Bigg]}{(k_1+k_2+k_3)^2+m^2} + \text{permutations}
\eea
where $s_{ij} =(k_i+k_j)^2$. 
Now putting in the expressions for the bare mass and bare coupling (Eqs.\ref{eq:Impm} and \ref{eq:Impc}) we get 
\bea 
&&G_{c}^{(6)}(k_1,...,k_6)\nonumber\\
&=& 
\chi_{4,\tilde p}^2\left(1-\frac{9}{N}\right)\frac{1}{(k_1+k_2+k_3)^2-\tilde p^2+\frac{1}{\chi_{2,\tilde p}}}  \nn\\
&+&  \chi_{4,\tilde p}^3\left(1 - \frac{12}{N}\right) \frac{\Bigg[V(s_{12})+V(s_{13})+V(s_{23})+V(s_{45})+V(s_{46})+V(s_{56})-2 V({\tilde s})-2V({\tilde t})-2V({\tilde  u})\Bigg]}{(k_1+k_2+k_3)^2-\tilde p^2+\frac{1}{\chi_{2,\tilde p}}} \nn\\
&-& \frac{6\chi_{4,\tilde p}^3}{N}\frac{\Bigg[V(s_{12})+V(s_{13})+V(s_{23})+V(s_{45})+V(s_{46})+V(s_{56})\Bigg]}{(k_1+k_2+k_3)^2-\tilde p^2+\frac{1}{\chi_{2,\tilde p}}} + \text{permutations}
\eea
The function defined in Eq.~\ref{eq:V} scales as $\ln \Lambda^2 \xi^2$ in 4D but is independent of the cut-off in lower dimensions.  Therefore in 4D, second line in the expression above is independent of the the UV cut-off, while the third line is sensitive to the cut-off and remains uncanceled. In 4D this gives a scaling $1/N \ln (\Lambda^2 \xi^2)$. Therefore at higher orders in $\lambda$ we can expect corrections of the form $1/N \left(\ln (\Lambda^2 \xi^2)\right)^n$. Hence the tuning of the coupling using the 4 point correlator does not guarantee cancellation of divergences that appear from the same types of corrections in higher point correlators. 

We therefore conclude that finite width (1/N) corrections are non-renormalizable. As we move to higher orders in  perturbation theory, as for the 2 and 4 point correlators, we will likewise have uncanceled 1/N contributions from powers of 1PI corrections to the external legs which will also be UV divergent. So the issue of non-renormalizability will persist even in lower dimensional theories that require renormalization. 

Strictly speaking, the breakdown of perturbation theory uncovered in this paper does not necessarily imply the non-existence of a non-perturbative definition of NNFT $\phi^4$ theory.Instead, we frame it as strongly indicative of the expected behavior in the  non-perturbative regime. The failure of perturbation theory stems from a combinatorial mismatch of lower point and higher point correlation functions which prevents perfect absorption of divergences. This mismatch appears to indicate that the original formulation will have uncontrolled errors when the cutoff is taken to infinity. We therefore interpret our results as limitation of the original formulation including the architectural modification proposed in this paper while leaving open the possibility that further improvements in the architecture may result in both a perturbative and non-perturbative definition of $\phi^4$ NNFT.
 
For the original proposed architecture, on face value it appears that to renormalize the theory, we would need to at least add an N dimensional local operator to absorb all the divergences for an N point correlator even for a one loop calculation. This is distinct, and much more severe than the for e.g., the non renormalizability of gravity, where we only need to add one higher dimensional operator for every loop order. However, we have seen that this problem could be solved by introducing functionals involving combinations of   activation function, distinct from usual higher dimensional local QFT counter-terms. So it may be possible obtain a better mapping of the target QFT to the the neural network by modifying the architecture.
%%%%%%%%%%%%%%%%%%%%%%%%%%%%%%%%%%%%%%%%%%%%%%%%%%%%%%%%%%%%%%%%%%%%%%%%%%%%%%%%%%%%%%%%%%%%%%%%5
\section{Summary and outlook}
\label{sec:Summary}

In this paper we took a detailed look at the perturbative implementation (perturbative in the coupling $\lambda$) of interacting NNFT at finite width using the proposed correspondence between NNFT and local QFTs in \cite{Demirtas:2023fir}. We work out in detail the one loop structure of this NNFT for the free and interacting scalar field theory. 
For the free field theory we observe that finite width corrections for a connected $n$ point correlator( $n \geq 4$) begins at O($1/N^{n/2-1}$) but more importantly, only gives contributions at isolated special kinematic points(SKPs) in momentum space and are zero otherwise. This means that as long as we keep away from these special kinematic points(SKP) in momentum space, there is no finite width error. 

For the interacting theory we consider a $\phi^4$ interaction and compute the two and four point connected correlators to one loop. We find three types of finite width contribution for the bare correlators at O(1/N): contribution proportional to the field theory result; contributions proportional to disconnected field theory diagrams;  non-gaussian contributions induced by the neural network. We perform perturbative renormalization tuning the bare mass and coupling to reproduce the two and four point observables working away from the SKP's. We observe that at one loop, the renormalized two point correlator only retains 1/N corrections from non-gaussian correlations that scale as $\Lambda^d\xi^d/N$ where $\Lambda$ is the momentum space cut-off and $\xi$ is the correlation length. All other types of 1/N corrections at one loop are absorbed in the bare parameters. 
The renormalized 4 point correlator at one loop retains three types of corrections at O(1/N); those proportional to the  field theory connected diagrams and suppressed by N; those from non-gaussian correlations that scale as $\Lambda^d\xi^d/N$; 1/N suppressed 1PI diagram corrections to external legs that scale( for $d >2$) as 1/N $\times \Lambda^{d-2}\xi^{d-2} $. While the first two are expected, the fact that the external leg corrections remain signals that the NNFT is not renormalizable and at O(n) in perturbation theory we expect corrections $1/N\left(\lambda \Lambda^{d-2}\xi^{d-2}\right)^n$, thus  restricting the convergence of the perturbative expansion to $\lambda \Lambda^{d-2}\xi^{d-2} \ll 1$. Likewise for accuracy of the result to O(n), we need $N \gg (\Lambda^{d-2}\xi^{d-2})^n$.  
To improve the convergence, we propose a modification to the probability distribution used for sampling the NN parameters which, among other 1/N corrections, also removes all  bubble diagrams from NNFT. While this will modify the renormalization group running of the mass, it improves the convergence of the perturbation theory now requiring $\lambda \Lambda \xi\ll 1$ in d=4 and $\lambda \ln \Lambda \xi\ll 1$ in d=3.

Finally we consider higher point correlators. We look at the corrections to  6 point connected correlator at one loop and show that even after tuning the bare mass and coupling using the two and four point correlators, the result for the renormalized six-point correlator contains UV divergent $1/N$ corrections proportional to QFT contribution.

The proposal architecture shows that the exact field theory results are reproduced in the $N \rightarrow \infty$. The expectation therefore was that the errors from having a finite width would scale as 1/N times the renormalized field theory result . However, we see that the scaling of the effects is much more severe and compared to a field theory result $\lambda^n$ and an expected error of $1/N\lambda^n$, the correction actually scales as  $1/N (\lambda \Lambda^{d-2} \xi^{d-2})^n$ for $d >2$ and $1/N (\lambda \ln \Lambda \xi)^n $ for d=2. With the improved perturbation theory framework, we have improved on this scaling as stated in the previous paragraph but this is still worse than the expected corrections $1/N \lambda^n $.

We conclude that using the current architecture, even with the improvement proposed in this paper, the finite width $1/N$ corrections in perturbation theory are non-renormalizable, and contain UV cut-off sensitive contributions that cannot be absorbed into the bare parameters of the theory. The improved architecture reduces the sensitivity to the cut-off but does not eliminate it as it reappears at higher orders and in corrections to higher point correlators. Throughout the draft, we used a hard cutoff on $b$ to compute observables and assess their dependence on divergences. The same can be repeated using softer regulators, e.g. an exponential or Gaussian damping of high $|b|$ modes. While we expect such damping to change the detailed form of the uncanceled divergent pieces that result from the combinatorial mismatch between diagrams, we expect the divergences to persist. This places constraints on the convergence of the perturbative expansion and the scaling for $N$ required to maintain accuracy of the result. 

All our analysis is based on the  NN architecture proposed in \cite{Demirtas:2023fir} and the modification we propose in this paper. However, it may be possible to  improve the utility of the neural network by further modifications to the architecture that would allows us to remove UV cut-off sensitive finite width corrections. New ideas are therefore needed to fully exploit the potential presented by this framework.

\section*{Acknowledgements}
V.V. would like to thank R.J. Furnstahl for introducing him to the field of NNFT. V.V. is supported by startup funds from the University of South Dakota and by the U.S. Department of Energy, EPSCoR program under contract No. DE-SC0025545. S.S. acknowledges support from the U.S.\ Department of Energy, Nuclear Physics Quantum Horizons program through the Early Career Award DE-SC0021892.

\appendix
%%%%%%%%%%%%%%%%%%%%%%%%%%%%%%%%%%%%%%%%%%%%%%%%%%%%%%%%%%%%%%%%%%%
\section*{Appendix}
Here we present technical details of some of the results on connected correlators in the free and interacting theory quoted in the main text. 
%%%%%%%%%%%%%%%%%%%%%%%%%%%%%%%%%%%%%%%%%%%%%%%%%%%%%%%%%%%%%%%%%%%%%%%%%%%%%%%%%%%
\section{Four point correlator in Free theory}
\label{App:4p}
We first look at free field theory at finite $N$. The calculation for the two-point correlator was presented in Section \ref{sec:2pFT}. We next consider the four- point correlator. The connected four-point correlator is defined as 
\bea
G_c^{(4)}(w_1,w_2,w_3,w_4) = G^{(4)}(w_1, w_2, w_3, w_4) 
- \Big\{G^{(2)}(w_1, w_2)  G^{(2)}(w_3, w_4) + \text{permutations}\Big\}.
\eea
Here 
\bea
&&G^{(4)}(w_1,w_2,w_3,w_4) \nonumber\\\nonumber\\
&\equiv&\langle \phi(w_1) \phi(w_2)\phi(w_3) \phi(w_4) \rangle    \nn\\\nonumber\\
&=&\left(\frac{2 V_d}{\sigma_a^2(2\pi)^d}\right)^2 \prod_{i=1}^N  \Bigg[ \int da_i  \frac{\sqrt{N}}{\sqrt{2\pi} \sigma_a} e^{-\frac{N}{2\sigma_a^2}a_ia_i}\int_{V_d} \frac{d^db_{i}}{V_d}\int_{-\pi}^{\pi} \frac{dc_i}{2\pi} \Bigg] \prod_{j=1}^4\sum_{k_j=1}^{N} \frac{a_{k_j} \cos (b_{k_j} \cdot w_j+c_{k_j})}{\sqrt{b_{k_j}^2+m^2}}.
\label{eq:four}
\eea
 
There are two types of contributions to the RHS of Eq. \ref{eq:four}, one is the disconnected contribution in which pairs of $a_{k_i}$ in the integrand are set equal to each other. We will call these pairwise contractions. The other is the connected contribution where all the $a_{k_i}'s$ are equal. We will call this four field/ non-Gaussian contraction. The latter is the deviation from Gaussianity. 
We can evaluate these two types of contractions separately 
\begin{enumerate}
\item{{\bf Pairwise contractions}:}
Pairwise contractions in Eq. \ref{eq:four} leads to 
\bea
G^{(4)}_{\text{p}}&=&\left(\frac{2 V_d}{\sigma_a^2(2\pi)^d}\right)^2 \prod_{i=1}^N  \Bigg[ \int da_i  \frac{\sqrt{N}}{\sqrt{2\pi} \sigma_a} e^{-\frac{N}{2\sigma_a^2}a_ia_i}\int_{V_d} \frac{d^db_{i}}{V_d}\int_{-\pi}^{\pi} \frac{dc_i}{2\pi} \Bigg] \times \left(\sum_{k_1=1}^{N} \frac{a^2_{k_1} \cos (b_{k_1} \cdot w_1+c_{k_1})\cos (b_{k_1} \cdot w_2+c_{k_1})}{b_{k_1}^2+m^2}\right)\nn\\
&&\left(\sum_{k_2\neq k_1}^{N} \frac{a^2_{k_2} \cos (b_{k_2} \cdot w_3+c_{k_2})\cos (b_{k_2} \cdot w_4+c_{k_2})}{b_{k_2}^2+m^2}\right) + w_2 \leftrightarrow w_3 + w_2 \leftrightarrow w_4
\eea
We first perform the integrals over $a_{k_1},a_{k_2}$ to get
\bea
G^{(4)}_{\text{p}}&=&\left(\frac{2 V_d}{N(2\pi)^d}\right)^2 \prod_{i=1}^N \Bigg[\int_{V_d} \frac{d^db_{i}}{V_d}\int_{-\pi}^{\pi} \frac{dc_i}{2\pi} \Bigg] \times \left(\sum_{k_1=1}^{N} \frac{ \cos (b_{k_1} \cdot w_1+c_{k_1})\cos (b_{k_1} \cdot w_2+c_{k_1})}{b_{k_1}^2+m^2}\right)\nn\\
&&\left(\sum_{k_2\neq k_1}^{N} \frac{\cos (b_{k_2} \cdot w_3+c_{k_2})\cos (b_{k_2} \cdot w_4+c_{k_2})}{b_{k_2}^2+m^2}\right)\nonumber\\
&&+ w_2 \leftrightarrow w_3 + w_2 \leftrightarrow w_4 .\nn
\eea
Next we do the integrals over $c_{k_1}$ and $c_{k_2}$ which gives
\bea
G^{(4)}_{\text{p}}&=&\left(\frac{V_d}{N(2\pi)^d}\right)^2 \prod_{i=1}^N \Bigg[\int_{V_d} \frac{d^db_{i}}{V_d}\Bigg] \times \left(\sum_{k_1=1}^{N} \frac{ \cos (b_{k_1} \cdot (w_1-w_2))}{b_{k_1}^2+m^2}\right)\left(\sum_{k_2\neq k_1}^{N} \frac{\cos (b_{k_2} \cdot (w_3-w_4))}{b_{k_2}^2+m^2}\right)\nonumber\\
&&+ w_2 \leftrightarrow w_3 + w_2 \leftrightarrow w_4. \nn
\eea
Now, we can do the integrals over $b_i$ giving us 
\bea 
G^{(4)}_{\text{p}}&=& \frac{N(N-1)}{N^2}\left(\int_{V_d} \frac{d^db_{1}}{(2\pi)^d} \frac{ \cos (b_{1} \cdot (w_1-w_2))}{b_{1}^2+m^2}\right)\left(\int_{V_d} \frac{d^db_{2}}{(2\pi)^d} \frac{\cos (b_{2} \cdot (w_3-w_4))}{b_{2}^2+m^2}\right)  + w_2 \leftrightarrow w_3 + w_2 \leftrightarrow w_4 \nn
\eea
which is proportional to the disconnected field theory diagrams. 
\item{{\bf Non- gaussian/ Four field contractions}}

 For this contribution, we set all four $a_{k_i}$ of the integrand to be equal giving us
 \bea
G^{(4)}_{\text{ng}}&=& \prod_{i=1}^N  \Bigg[ \int da_i  \frac{\sqrt{N}}{\sqrt{2\pi} \sigma_a} e^{-\frac{N}{2\sigma_a^2}a_ia_i}\int_{V_d} \frac{d^db_{i}}{V_d}\int_{-\pi}^{\pi} \frac{dc_i}{2\pi} \Bigg]\nn\\
&\times & \left(\frac{2 V_d}{\sigma_a^2(2\pi)^d}\right)^2\sum_{k=1}^{N} \frac{a^4_{k} \cos (b_{k} \cdot w_1+c_{k})\cos (b_{k} \cdot w_2+c_{k})\cos(b_k\cdot w_3+c_{k})\cos (b_{k} \cdot w_4+c_{k})}{(b_{k}^2+m^2)^2}.
\eea
As before we first do the integral over $a_{k_i}$ to get
\bea
&&G^{(4)}_{\text{ng}}\nonumber\\
&=& 3\left(\frac{2 V_d}{N(2\pi)^d}\right)^2\prod_{i=1}^N \Bigg[\int_{V_d} \frac{d^db_{i}}{V_d}\int_{-\pi}^{\pi} \frac{dc_i}{2\pi} \Bigg] \sum_{k=1}^{N} \frac{ \cos (b_{k} \cdot w_1+c_{k})\cos (b_{k} \cdot w_2+c_{k})\cos(b_k\cdot w_3+c_{k})\cos (b_{k} \cdot w_4+c_{k})}{(b_{k}^2+m^2)^2}\nn
\eea
followed by the integral over $c_{k}$ to get 
\small
\bea
G^{(4)}_{\text{ng}}&=& \frac{3}{2}\left(\frac{V_d}{N(2\pi)^d}\right)^2\prod_{i=1}^N \Bigg[\int_{V_d} \frac{d^db_{i}}{V_d} \Bigg] \nonumber\\
&&\times \sum_{k=1}^{N} \frac{ \cos (b_{k} \cdot (w_1+w_2-w_3-w_4)+\cos (b_{k} \cdot (w_1-w_2+w_3-w_4))+\cos(b_k\cdot (w_1-w_2-w_3+w_4))}{(b_{k}^2+m^2)^2}.\nn
\eea
\normalsize

Performing the $b_i$ integrals we get 
\bea
G^{(4)}_{\text{ng}}&=& \frac{3V_d}{2N(2\pi)^d}\int_{V_d} \frac{d^db}{(2\pi)^d}\frac{e^{ib\cdot (w_1+w_2-w_3-w_4)}+e^{ib \cdot (w_1-w_2+w_3-w_4)}+e^{i b\cdot (w_1-w_2-w_3+w_4)}}{(b^2+m^2)^2}\nn
\eea
which explicitly scales as $1/N$. We can get the four point correlator as 
\bea
G^{(4)}=
G^{(4)}_{\text{p}}+G^{(4)}_{\text{ng}}.\eea

The connected 4 point correlator is then given by
\end{enumerate}
\bea
 &&G_{c}^{(4)}(w_1,w_2,w_3,w_4)= \frac{3V_d}{2N(2\pi)^d}\int_{V_d} \frac{d^db}{(2\pi)^d}\frac{e^{ib \cdot (w_1+w_2-w_3-w_4)}+e^{ib \cdot (w_1-w_2+w_3-w_4)}+e^{i b\cdot (w_1-w_2-w_3+w_4)}}{(b^2+m^2)^2}\nn\\
&-& \Bigg\{\frac{1}{N} \left(\int_{V_d} \frac{d^db_{1}}{(2\pi)^d} \frac{ e^{i b_{1} \cdot (w_1-w_2)}}{b_{1}^2+m^2}\right)\left(\int_{V_d} \frac{d^db_{2}}{(2\pi)^d} \frac{e^{ib_{2} \cdot (w_3-w_4)}}{b_{2}^2+m^2}\right)  + w_2 \leftrightarrow w_3 + w_2 \leftrightarrow w_4 \Bigg\}.
\eea

%%%%%%%%%%%%%%%%%%%%%%%%%%%%%%%%%%%%%%%%%%%%%%%%%%%%%%%%%%%%%%%%%%%%%%%%%%%%%%%%%%%%%%%%%%%%%%%%%%%%%%%%%%%%%%%%%%%%%

\section{Two point correlator in interacting theory}
\label{App:twoP}
Here we look at the computation of connected correlators in the interacting theory to one loop. We will only discuss the two point correlator in detail since the computation of all higher point correlators, while more elaborate, follows the same procedure. The two point correlator in interacting theory  to O($\lambda$) was defined in Eq.~\ref{eq:2pIT} which we write here for convenience
\bea
\langle \phi(w_1) \phi(w_2) \rangle = \langle \phi(w_1)\phi(w_2) \rangle_{\text{f}}-\frac{\lambda}{4!}\int d^dx \langle \phi(w_1)\phi(w_2) \phi^4(x)\rangle_{\text{f}}+\langle \phi(w_1) \phi(w_2)\rangle_{\text{f}}\frac{\lambda}{4!}\int d^dx \langle \phi^4(x) \rangle_{\text{f}} 
\label{eq:2pNLO}.
\eea
We now consider each term at O($\lambda$) in turn. 

\begin{enumerate}
    \item{} 
    \bea
  &&  \frac{\lambda}{4!}\int d^dx \langle \phi(w_1)\phi(w_2) \phi^4(x)\rangle_{\text{f}}  = 
     \prod_{i=1}^N  \Bigg[ \int da_i  \frac{\sqrt{N}}{\sqrt{2\pi} \sigma_a} e^{-\frac{N}{2\sigma_a^2}a_ia_i}\Bigg]\Bigg[ \frac{1}{V_d}\int_{V_d} d^db_{i}  \Bigg] \Bigg[\frac{1}{2\pi} \int_{-\pi}^{\pi} dc_i \Bigg] \left(\frac{2 V_d}{\sigma_a^2(2\pi)^d}\right)^3 \nn\\ 
&&\prod_{j=1}^2\left( \sum_{k_j}^{N} \frac{a_{k_j} \cos (b_{k_j}\cdot  w_j+c_{k_j})}{\sqrt{b_{k_j}^2+m^2}}\right) \frac{\lambda}{4!}\int d^dx  \prod_{l=1}^4\left( \sum_{u_l=1}^{N} \frac{a_{u_l} \cos (b_{u_l} \cdot x+c_{u_l})}{\sqrt{b_{u_l}^2+m^2}}\right) 
    \eea
    From our calculation for the free theory we again identify two types of corrections, pairwise contractions that lead to contributions proportional to field theory diagrams and non-gaussian four field contractions.
\begin{itemize}
    \item {{\bf Pairwise contractions}:}
    
    This will give two types of corrections corresponding to the diagrams shown in Fig.~\ref{fig:2pD}. 
    \begin{figure}
\centering
\includegraphics[width=0.5\linewidth]{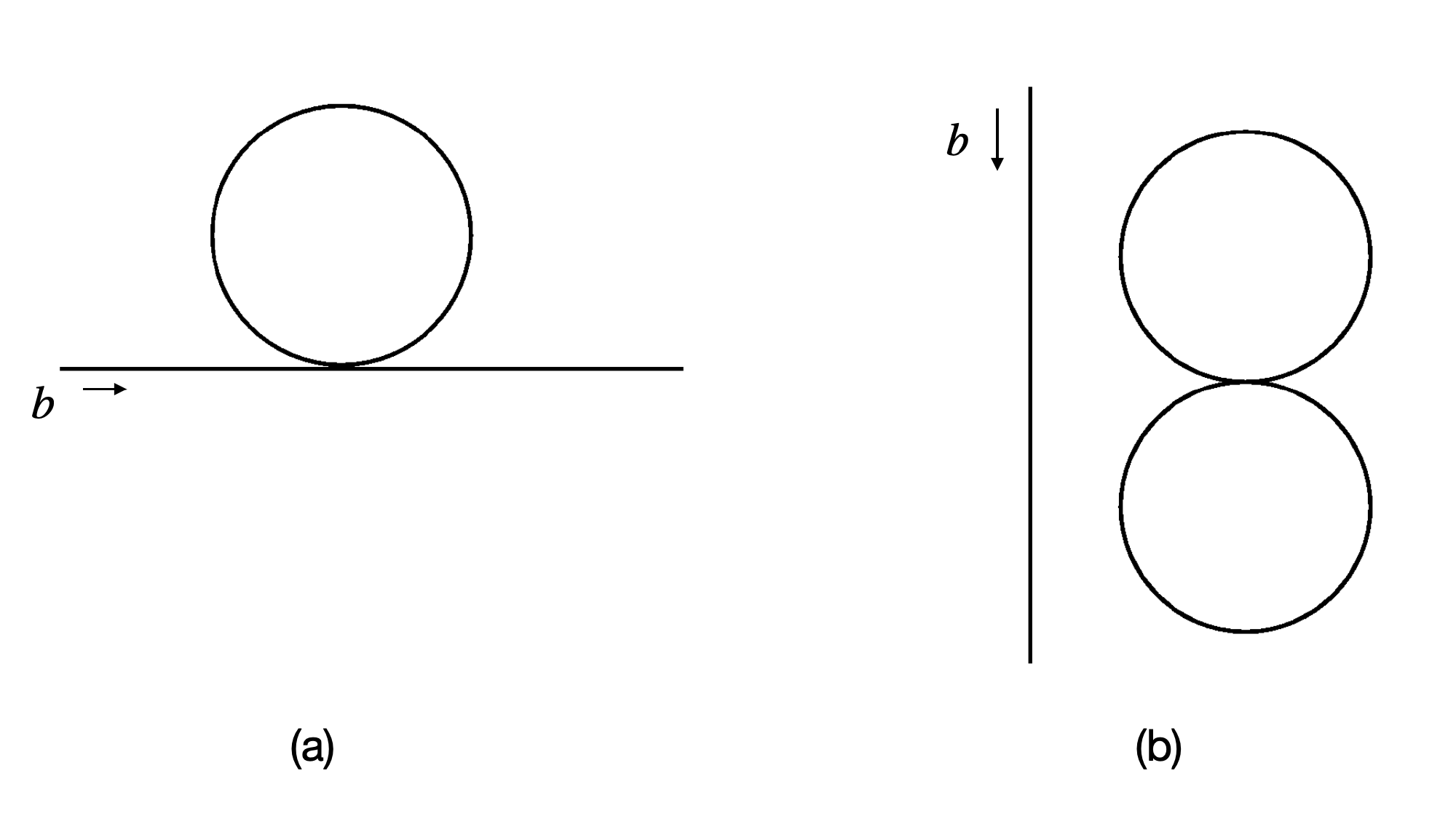}
\label{fig:2pD}
\caption{Connected and disconnected Feynman diagrams for two point correlator.}
\end{figure}
\bea
  &&  \frac{\lambda}{4!}\int d^dx \langle \phi(w_1)\phi(w_2) \phi^4(x)\rangle_{\text{\text{f,p}}}\nonumber\\
  &=& 
     \prod_{i=1}^N  \Bigg[ \int da_i  \frac{\sqrt{N}}{\sqrt{2\pi} \sigma_a} e^{-\frac{N}{2\sigma_a^2}a_ia_i}\Bigg]\Bigg[ \frac{1}{V_d}\int_{V_d} d^db_{i}  \Bigg] \Bigg[\frac{1}{2\pi} \int_{-\pi}^{\pi} dc_i \Bigg] \left(\frac{2 V_d}{\sigma_a^2(2\pi)^d}\right)^3 \nn\\ 
&&\int d^dx \Bigg[ \frac{\lambda}{2}\prod_{j=1}^2\left( \sum_{k_j, k_1 \neq k_2}^{N} \frac{a^2_{k_j} \cos (b_{k_j}\cdot  w_j+c_{k_j}) \cos (b_{k_j}\cdot  x+c_{k_j})}{b_{k_j}^2+m^2}\right) \sum_{u \neq k_1, k_2}^{N} \left(\frac{a_{u} \cos (b_{u} \cdot x+c_{u})}{\sqrt{b_{u}^2+m^2}}\right)^2 \nn\\
&+ & \frac{\lambda}{8}\prod_{j=1}^2\sum_{u_j, u_1 \neq u_2}^{N} \left(\frac{a_{u_j} \cos (b_{u_j}\cdot  x+c_{u_j})}{b_{u_j}^2+m^2}\right)^2 \sum_{ k\neq u_1,u_2}^{N} \left(\frac{a^2_{k} \cos (b_{k} \cdot w_1 +c_{k})\cos (b_{k} \cdot w_2 +c_{k})}{b_{k}^2+m^2}\right) \Bigg].
\eea
We can now evaluate the integrals over $a_i$ followed by $c_i$ followed by $b_i$ leaving us with 
\bea
  &&  \frac{\lambda}{4!}\int d^dx \langle \phi(w_1)\phi(w_2) \phi^4(x)\rangle_{\text{\text{f,p}}}
  \nonumber\\
  &=& \frac{N(N-1)(N-2)}{N^3}\Bigg[ \frac{\lambda}{2}\int_{V_d}\frac{d^db}{(2\pi)^d}\frac{e^{ib (w_1-w_2)}}{(b^2+m^2)^2}\int_{V_d} \frac{d^dq}{(2\pi)^d}\frac{1}{q^2+m^2}  \nn\\
  &+&\frac{\lambda}{8}\int_{V_d}\frac{d^db}{(2\pi)^d}\frac{e^{ib (w_1-w_2)}\delta^d(0)}{b^2+m^2}\left(\int_{V_d} \frac{d^dq}{(2\pi)^d}\frac{1}{q^2+m^2} \right)^2\Bigg]
\eea
\item{{\bf Non Gaussian /four field contractions}:}

This will give three types of contributions corresponding to the 3 distinct ways in which 4 fields can be grouped leading to 
\bea
 &&  \frac{\lambda}{4!}\int d^dx \langle \phi(w_1)\phi(w_2) \phi^4(x)\rangle_{\text{f,ng}}  = 
     \prod_{i=1}^N  \Bigg[ \int da_i  \frac{\sqrt{N}}{\sqrt{2\pi} \sigma_a} e^{-\frac{N}{2\sigma_a^2}a_ia_i}\Bigg]\Bigg[ \frac{1}{V_d}\int_{V_d} d^db_{i}  \Bigg] \Bigg[\frac{1}{2\pi} \int_{-\pi}^{\pi} dc_i \Bigg] \left(\frac{2 V_d}{\sigma_a^2(2\pi)^d}\right)^3 \nn\\ 
&&\int d^dx \Bigg[\frac{\lambda}{4}\sum_{k=1}^{N} \frac{a^4_{k} \cos (b_{k}\cdot  w_1+c_{k}) \cos (b_{k}\cdot  w_2+c_{k})\cos^2(b_{k}\cdot x+c_{k})}{(b_{k}^2+m^2)^2}\sum_{u \neq k}^{N} \left(\frac{a_{u} \cos (b_{u} \cdot x+c_{u})}{\sqrt{b_{u}^2+m^2}}\right)^2 \nn\\
&+&\Bigg\{\frac{\lambda}{6}\sum_{k=1}^{N} \frac{a^4_{k} \cos (b_{k}\cdot  w_1+c_{k}) \cos^3(b_{k}\cdot x+c_{k})}{(b_{k}^2+m^2)^2}\sum_{u \neq k}^{N} \frac{a^2_{u} \cos (b_{u} \cdot x+c_{u})\cos (b_{u}\cdot  w_2+c_{u})}{b_{u}^2+m^2} + w_1 \leftrightarrow w_2\Bigg\}\nn\\
&+&\frac{\lambda}{4!}\sum_{k=1}^{N} \frac{a^4_{k}\cos^4(b_{k}\cdot x+c_{k})}{(b_{k}^2+m^2)^2}\sum_{u \neq k}^{N} \frac{a^2_{u} \cos (b_{u}\cdot  w_1+c_{u}) \cos (b_{u}\cdot  w_2+c_{u})}{b_{u}^2+m^2} \Bigg].
\eea
We only consider grouping of up to 4 fields in this paper since we are interested in the O(1/N) corrections. 
Grouping more than 4 fields will yield higher powers of 1/N. 
We now evaluate the $a_i,b_i$ and $c_i$ integrals to get
\bea
 &&  \frac{\lambda}{4!}\int d^dx \langle \phi(w_1)\phi(w_2) \phi^4(x)\rangle_{\text{f,ng}}  = 
      \frac{ V_d}{(2\pi)^d}\frac{N(N-1)}{N^3}\int_{V_d}\frac{d^db}{(2\pi)^d}e^{i b(w_1-w_2)}\nn\\ 
&&\Bigg[\frac{3\lambda}{8}\frac{(2\delta^d(0)+ \delta^d(2b))}{(b^2+m^2)^2}\int_{V_d}\frac{d^dq}{(2\pi)^d}\frac{1}{q^2+m^2} +\frac{3\lambda}{2}\frac{1}{(b^2+m^2)^3} +\frac{3\lambda}{16} \frac{\delta^d(0)}{b^2+m^2}\int_{V_d}\frac{d^dq}{(2\pi)^d}\frac{1}{(q^2+m^2)^2}\Bigg].
\eea
\end{itemize}

\item{}
Next we consider the third term in Eq.~\ref{eq:2pNLO}. We usually ignore this term while computing Feynman diagrams in field theory since they only lead to disconnected contributions which cancel out with the corresponding diagrams obtained from the second term of Eq.~\ref{eq:2pNLO}. However we expect that this will not hold for the O(1/N) corrections 
\bea
\langle \phi(w_1) \phi(w_2)\rangle_{\text{f}}\frac{\lambda}{4!}\int d^dx \langle \phi^4(x) \rangle_{\text{f}} = \int_{V_d}\frac{d^db}{(2\pi)^d}\frac{e^{ib\cdot(w_1-w_2)}}{b^2+m^2}\times \frac{\lambda}{4!}\int d^dx \langle \phi^4(x) \rangle_{\text{f}}.
\eea
We can focus on the piece $\frac{\lambda}{4!}\int d^dx \langle \phi^4(x) \rangle_{\text{f}}$ where we have
\begin{itemize}
    \item{{\bf Pairwise contractions}:}
    \bea 
    &&\frac{\lambda}{4!}\int d^dx \langle \phi^4(x) \rangle_{\text{\text{f,p}}} =  \prod_{i=1}^N  \Bigg[ \int da_i  \frac{\sqrt{N}}{\sqrt{2\pi} \sigma_a} e^{-\frac{N}{2\sigma_a^2}a_ia_i}\Bigg]\Bigg[ \frac{1}{V_d}\int_{V_d} d^db_{i}  \Bigg] \Bigg[\frac{1}{2\pi} \int_{-\pi}^{\pi} dc_i \Bigg] \left(\frac{2 V_d}{\sigma_a^2(2\pi)^d}\right)^2 \nn\\ 
    &&\int d^dx \frac{\lambda}{8} \sum_{k}^{N} \left(\frac{a_{k} \cos (b_{k} \cdot x+c_{k})}{\sqrt{b_{k}^2+m^2}}\right)^2\sum_{u \neq k}^{N} \left(\frac{a_{u} \cos (b_{u} \cdot x+c_{u})}{\sqrt{b_{u}^2+m^2}}\right)^2 =\frac{N(N-1)}{N^2} \frac{\lambda}{8} \delta^d(0) \left(\int_{V_d} \frac{d^dq}{(2\pi)^d}\frac{1}{q^2+m^2} \right)^2\nonumber\\
    \eea
    and 
    \item{{\bf Non Gaussian/Four field contractions}:}
 \bea 
   && \frac{\lambda}{4!}\int d^dx \langle \phi^4(x) \rangle_{\text{f,ng}} =  \prod_{i=1}^N  \Bigg[ \int da_i  \frac{\sqrt{N}}{\sqrt{2\pi} \sigma_a} e^{-\frac{N}{2\sigma_a^2}a_ia_i}\Bigg]\Bigg[ \frac{1}{V_d}\int_{V_d} d^db_{i}  \Bigg] \Bigg[\frac{1}{2\pi} \int_{-\pi}^{\pi} dc_i \Bigg] \left(\frac{2 V_d}{\sigma_a^2(2\pi)^d}\right)^2 \nn\\ &&\frac{\lambda}{4!}\int d^dx  \sum_{k=1}^{N} \frac{a^4_{k}\cos^4(b_{k}\cdot x+c_{k})}{(b_{k}^2+m^2)^2} = \frac{ V_d}{(2\pi)^d}\frac{1}{N}\frac{3\lambda}{16} \delta^d(0)\int_{V_d}\frac{d^dq}{(2\pi)^d}\frac{1}{(q^2+m^2)^2}.
    \eea
    
\end{itemize}

\end{enumerate}
Combining all of the contributions at O($\lambda$) and only retaining up to O$(1/N)$ corrections we obtain Eq.~\ref{eq:2p}.

\bibliography{AI.bib}
\end{document}